\documentclass[apj]{emulateapj}
\usepackage{apjfonts}
\usepackage{amsmath}
\usepackage{enumerate}
\usepackage{natbib}

\newcommand{\pivec}{\mbox{\boldmath $\pi$}}

\shorttitle{The 2L1S/1L2S degeneracy}
\shortauthors{Shin et al.} 

\begin{document}

\title{
The 2L1S/1L2S Degeneracy for Two Microlensing Planet Candidates Discovered by the KMTNet Survey in 2017
}

\author{
I.-G.~Shin\altaffilmark{K1,H1},
J.~C.~Yee\altaffilmark{H1},
A.~Gould\altaffilmark{K1,E1,E2},
M.~T.~Penny\altaffilmark{E1},
I.~A.~Bond\altaffilmark{M1},\\ 
and\\
M.~D.~Albrow\altaffilmark{K3}, 
S.-J.~Chung\altaffilmark{K1,K2},
C.~Han\altaffilmark{K5},
K.-H.~Hwang\altaffilmark{K1},
Y.~K.~Jung\altaffilmark{K1},
Y.-H.~Ryu\altaffilmark{K1},
Y.~Shvartzvald\altaffilmark{E3},
S.-M.~Cha\altaffilmark{K1,K4},   
D.-J.~Kim\altaffilmark{K1},
H.-W.~Kim\altaffilmark{K1,K2},
S.-L.~Kim\altaffilmark{K1,K2},
C.-U.~Lee\altaffilmark{K1,K2},  
D.-J.~Lee\altaffilmark{K1},
Y.~Lee\altaffilmark{K1,K4},
B.-G.~Park\altaffilmark{K1,K2},
R.~W.~Pogge\altaffilmark{E1}\\
(KMTNet Collaboration),\\
F.~Abe\altaffilmark{M2}, 
R.~Barry\altaffilmark{M3},
D.~P.~Bennett\altaffilmark{M3}
A.~Bhattacharya\altaffilmark{M3,M4}, 
M.~Donachie\altaffilmark{M5}, 
H.~Fujii\altaffilmark{M2},
A.~Fukui\altaffilmark{M6}, 
Y.~Hirao\altaffilmark{M7},
Y.~Itow\altaffilmark{M2}, 
Y.~Kamei\altaffilmark{M2},
Iona~Kondo\altaffilmark{M7},
N.~Koshimoto\altaffilmark{M7}, 
M.~C.~A.~Li\altaffilmark{M5}, 
Y.~Matsubara\altaffilmark{M2}, 
S.~Miyazaki\altaffilmark{M7}, 
Y.~Muraki\altaffilmark{M2},
M.~Nagakane\altaffilmark{M7}, 
C.~Ranc\altaffilmark{M3,F1}, 
N.~J.~Rattenbury\altaffilmark{M5}, 
Harmon~Suematsu\altaffilmark{M7},
D.~J.~Sullivan\altaffilmark{M8},
T.~Sumi\altaffilmark{M7},
Daisuke~Suzuki\altaffilmark{M9},
P.~J.~Tristram\altaffilmark{M10}, 
T.~Yamakawa\altaffilmark{M2},
A.~Yonehara\altaffilmark{M11}\\
(MOA Collaboration),\\
and\\
P.~Fouqu{\'e}\altaffilmark{E5,E6},
W.~Zang\altaffilmark{E4}\\
(CFHT-K2C9 Microlensing Collaboration)\\
}

\bigskip\bigskip
\affil{$^{K1}$Korea Astronomy and Space Science Institute, 776 Daedeokdae-ro, Yuseong-Gu, Daejeon 34055, Republic of Korea}
\affil{$^{H1}$Harvard-Smithsonian Center for Astrophysics, 60 Garden St., Cambridge, MA 02138, USA}
\affil{$^{K2}$Korea University of Science and Technology, 217 Gajeong-ro, Yuseong-gu, Daejeon 34113, Republic of Korea}
\affil{$^{K3}$University of Canterbury, Department of Physics and Astronomy, Private Bag 4800, Christchurch 8020, New Zealand}
\affil{$^{K4}$School of Space Research, Kyung Hee University, Giheung-gu, Yongin, Gyeonggi-do, 17104, Republic of Korea}
\affil{$^{K5}$Department of Physics, Chungbuk National University, Cheongju 28644, Republic of Korea}
\affil{$^{M1}$Institute of Natural and Mathematical Sciences, Massey University, Auckland 0745, New Zealand}
\affil{$^{M2}$Institute for Space-Earth Environmental Research, Nagoya University, Nagoya 464-8601, Japan}
\affil{$^{M3}$Code 667, NASA Goddard Space Flight Center, Greenbelt, MD} 
\affil{$^{M4}$Department of Astronomy, University of Maryland, College Park, MD, USA}
\affil{$^{M5}$Department of Physics, University of Auckland, Private Bag 92019, Auckland, New Zealand}
\affil{$^{M6}$Okayama Astrophysical Observatory, National Astronomical Observatory of Japan, 3037-5 Honjo, Kamogata, Asakuchi, Okayama 719-0232, Japan}
\affil{$^{M7}$Department of Earth and Space Science, Graduate School of Science, Osaka University, Toyonaka, Osaka 560-0043, Japan}
\affil{$^{M8}$School of Chemical and Physical Sciences, Victoria University, Wellington, New Zealand}
\affil{$^{M9}$Institute of Space and Astronautical Science, Japan Aerospace Exploration Agency, Kanagawa 252-5210, Japan}
\affil{$^{M10}$University of Canterbury Mount John Observatory, P.O. Box 56, Lake Tekapo 8770, New Zealand}
\affil{$^{M11}$Department of Physics, Faculty of Science, Kyoto Sangyo University, 603-8555 Kyoto, Japan}
\affil{$^{E1}$Department of Astronomy, Ohio State University, 140 W. 18th Ave., Columbus, OH 43210, USA}
\affil{$^{E2}$Max-Planck-Institute for Astronomy, K\"onigstuhl 17, 69117 Heidelberg, Germany}
\affil{$^{E3}$IPAC, Mail Code 100-22, Caltech, 1200 E. California Blvd., Pasadena, CA 91125, USA}
\affil{$^{E4}$Physics Department and Tsinghua Centre for Astrophysics, Tsinghua University, Beijing 100084, People's Republic of China}
\affil{$^{E5}$CFHT Corporation, 65-1238 Mamalahoa Hwy, Kamuela, Hawaii 96743, USA}
\affil{$^{E6}$Universit{\'e} de Toulouse, UPS-OMP, IRAP, Toulouse, France}
\affil{$^{F1}$NASA Postdoctoral Program Fellow}

\begin{abstract}
 We report two microlensing planet candidates discovered by the KMTNet survey in $2017$. However, both events have the 2L1S/1L2S degeneracy, which is an obstacle to claiming the discovery of the planets with certainty unless the degeneracy can be resolved. For KMT-2017-BLG-0962, the degeneracy cannot be resolved. If the 2L1S solution is correct, KMT-2017-BLG-0962 might be produced by a super Jupiter-mass planet orbiting a mid-M dwarf host star. For KMT-2017-BLG-1119, the light curve modeling favors the 2L1S solution but higher-resolution observations of the baseline object tend to support the 1L2S interpretation rather than the planetary interpretation. This degeneracy might be resolved by a future measurement of the lens-source relative proper motion. This study shows the problem of resolving 2L1S/1L2S degeneracy exists over a much wider range of conditions than those considered by the theoretical study of \citet{gaudi98}. 
\end{abstract}

\section{Introduction}
 The basic requirements for the statistical studies of planets are detections of planets and the determination of planet properties. However, discoveries and characterizations of microlensing planets depend on the interpretation of anomalies in the observed light curves. Even when these anomalies can be described by a planetary model, there may exist alternative interpretations that also provide sufficient descriptions for the putative planetary anomalies. In other words, degenerate solutions of the light curves can be obstacles to prevent either secure discoveries of planets or the unique determination of their properties. 

 For example, the degeneracy between two interpretations of the binary-lens and single-source (2L1S) and the single-lens and binary-source (1L2S) can be a severe obstacle. If this 2L1S/1L2S degeneracy exists, we cannot claim a secure discovery of the planet unless the degeneracy is resolved. \citet{gaudi98} first pointed out this 2L1S/1L2S degeneracy by showing that a certain class of 1L2S model can resemble a planetary anomaly in the lensing light curve. In particular, he focused on planetary events that exhibit small, short-duration positive deviations from a single-lens, single-source (1L1S) light curve. To produce a similar anomaly in the light curve using a 1L2S model, the brightness of the companion should be much fainter than the primary (the flux ratio of the secondary and primary, $\epsilon \equiv F_{2}/F_{1}$, should be from $\epsilon \sim 10^{-2}$ to $\sim 10^{-4}$). In addition, the companion should pass very close (in projection) to the lens (this impact factor for the secondary, $u_{0,S2}$, depends on the  maximum amplitude, $\delta_{\rm max}$, of the planet-like anomaly with the flux ratio: $u_{0,S2} \lesssim \epsilon / {\delta_{\rm max}}$).

 Indeed, there are discoveries of microlensing planet candidates, which could be interpreted by both 2L1S and 1L2S models. \citet{beaulieu06} found a clear planetary deviation (i.e., a small, short-duration positive deviation) in a microlensing event, OGLE-2005-BLG-390. They also found the 2L1S/1L2S degeneracy that plausibly described the anomaly. However, the 1L2S interpretation was rejected by the detailed light curve analysis. Thus, they could claim the secure discovery of a planet, whose mass they estimated to be $5.5\,M_{\oplus}$. \citet{hwang13} also showed a microlensing event that had the 2L1S/1L2S degeneracy. The light curve of this work exhibits a planet-like anomaly (i.e., the strong positive deviation) that can be explained by either the 2L1S (including a planet) or 1L2S interpretations. They successfully resolved this degeneracy using multi-band observations revealing that the event was produced by two sources, rather than a planetary system. In addition, \citet{dominik19} recently presented a long timescale ($t_{\rm E}\sim 300$ days) microlensing event, which can be explained either 2L1S or 1L2S interpretations. Their 2L1S model indicates that the lens system might be a planet with the mass $\sim45\, M_{\oplus}$ orbiting an M-dwarf host star ($\sim0.35\, M_{\odot}$). However, they also find a competitive 1L2S model that indicates that the lens might be a brown-dwarf ($0.046\, M_{\odot}$). The light curve data cannot resolve this degeneracy, but they suggest future observations may be able to resolve this severe degeneracy.

 However, in practice, we have found that the 2L1S/1L2S degeneracy can be extended to cases beyond the extreme flux case considered by \citet{gaudi98}, e.g., \citet{jung17a}, \citet{dominik19}, and events in this work. In \citet{jung17a}, the light curve of the event showed a broad asymmetry with small additional deviations in the wing. This anomaly can be adequately described by both the 2L1S (i.e., a planetary lens system) and the 1L2S interpretations. This event was produced by close to equal-luminous binary sources in contrast to the case of \citet{gaudi98}. They resolved this degeneracy using detailed modeling of the densely covered light curve. In \citet{dominik19}, they showed that the planet-like anomaly in the 2L1S case could be produced when the source passes close to the central caustic, i.e., a high-magnification event. This anomaly is different from Gaudi's case, which is produced when the source approaches one of planetary caustics. They noted that the 1L2S model with a small flux ratio of binary sources can produce this planet-like anomaly in contrast to Gaudi's case. 

 In addition, microlensing events showing more complex anomalies have been found. These events can be described by more complicated multiple-lens and multiple-source interpretations. For example, \citet{jung17b} showed a degeneracy caused by 3L1S and 2L2S interpretations. Moreover, \citet{hwang18} showed an extreme case (i.e., exo-moon candidate) of a three-fold degeneracy with 3L1S, 2L2S, and 1L3S interpretations. In particular, the degeneracy becomes severe when the observations do not optimally cover the anomalies in the light curves.

 Here we analyze two microlensing events, KMT-2017-BLG-0962 and KMT-2017-BLG-1119, that were discovered in $2017$ by the Korea Microlensing Telescope Network \citep[KMTNet:][]{kim16}. We reveal that these events are planet candidates by analyzing the light curves using the 2L1S interpretation. For KMT-2017-BLG-0962, the mass ratio ($q=M_{\rm planet}/M_{\rm host}$) is $\sim 0.01$, which indicates that the companion in the lens system might be a Jupiter-class planet under the assumption of an M-dwarf host star. For KMT-2017-BLG-1119, the mass ratio is $\sim 0.01$, which also indicates that the lens component might be a planet. Moreover, the Einstein timescale ($t_{\rm E}$) of this event is very short, i.e., $t_{\rm E}\sim 2.9$ days. This short timescale implies that the event can be produced by a very low-mass planetary lens system\footnote[1]{The Einstein timescale is a crossing time that the source transverses the Einstein ring radius ($\theta_{\rm E}$), i.e., $t_{\rm E} \propto \theta_{\rm E}$. The size of $\theta_{\rm E}$ is directly related to the mass of the lens system ($M$), i.e., $\theta_{\rm E} \propto \sqrt{M / D_{\rm rel}}$ where $D_{\rm rel} \equiv (D_{\rm L}^{-1} - D_{\rm S}^{-1})^{-1}$. $D_{\rm L}$ and $D_{\rm S}$ are distances to the lens and source, respectively. Thus, $t_{\rm E} \propto \sqrt{M}$, which are of order a month for typical microlensing events.}. However, both light curves can also be well described using the 1L2S interpretation. 

 We present observations of these planet candidates in Section 2. In Section 3, we present analyses of the light curves and the degeneracies. Then, we discuss the possibilities of resolving the degeneracies in Section 4. In Section 5, we present the possible properties of planet candidates determined using the Bayesian analyses. Lastly, in Section 6, we present our conclusion with the difference between a \citet{gaudi98}-type degeneracy and this work. Additionally, we provide details of the 1L2S interpretations for the modeling in Appendix A. We also present tests for higher-order effects of the models to discuss non-detections of them in Appendix B.

\section{KMTNet Observations}
 KMTNet is a second-generation microlensing survey consisting of a telescope network composed of three identical $1.6$ m telescopes located at three sites in the southern hemisphere: the Cerro Tololo Inter-American Observatory in Chile (KMTC), the South African Astronomical Observatory in South Africa (KMTS), and the Siding Spring Observatory in Australia (KMTA). These well-separated time zones can provide near-continuous observations, weather permitting. In addition, the cameras of the KMTNet survey have a wide field of view (FOV: $4$ ${\rm deg}^2$). These wide FOV yield high-cadence observations that are optimized to capture planetary anomalies caused by various types of planets. Thus, in general, the KMTNet survey (i.e., a second-generation microlensing survey) is less dependent on follow-up observations. 

 KMTNet discovered the two planet candidates presented in this work. The events were found by the KMTNet Event Finder algorithm \citep{kim18}, which was run after the end of the $2017$ microlensing season. No real-time alert was issued for these events, either by KMTNet or other microlensing groups. Hence, no useful real-time photometric follow-up observations were taken\footnote[2]{KMT-2017-BLG-1119 was in fact serendipitously observed by the {\it Spitzer} satellite because it lies within the IRAC camera field of view of another event (OGLE-2017-BLG-0019) that was chosen for observations (see \citealt{yee15}). Unfortunately, these observations ended (due to sun-angle restrictions) on ${\rm JD}-2450000.0\sim7967.0$, just two days before the peak of this very short event. In principle, if the lens were traveling approximately east, the source could nevertheless have been significantly magnified.  However, we have checked the images and found that the {\it Spitzer} light curve of KMT-2017-BLG-1119 is essentially flat.  Thus, no meaningful constraints can be placed on this system from the {\it Spitzer} data.}.

\begin{figure*}[htb!]
\epsscale{1.00}
\plotone{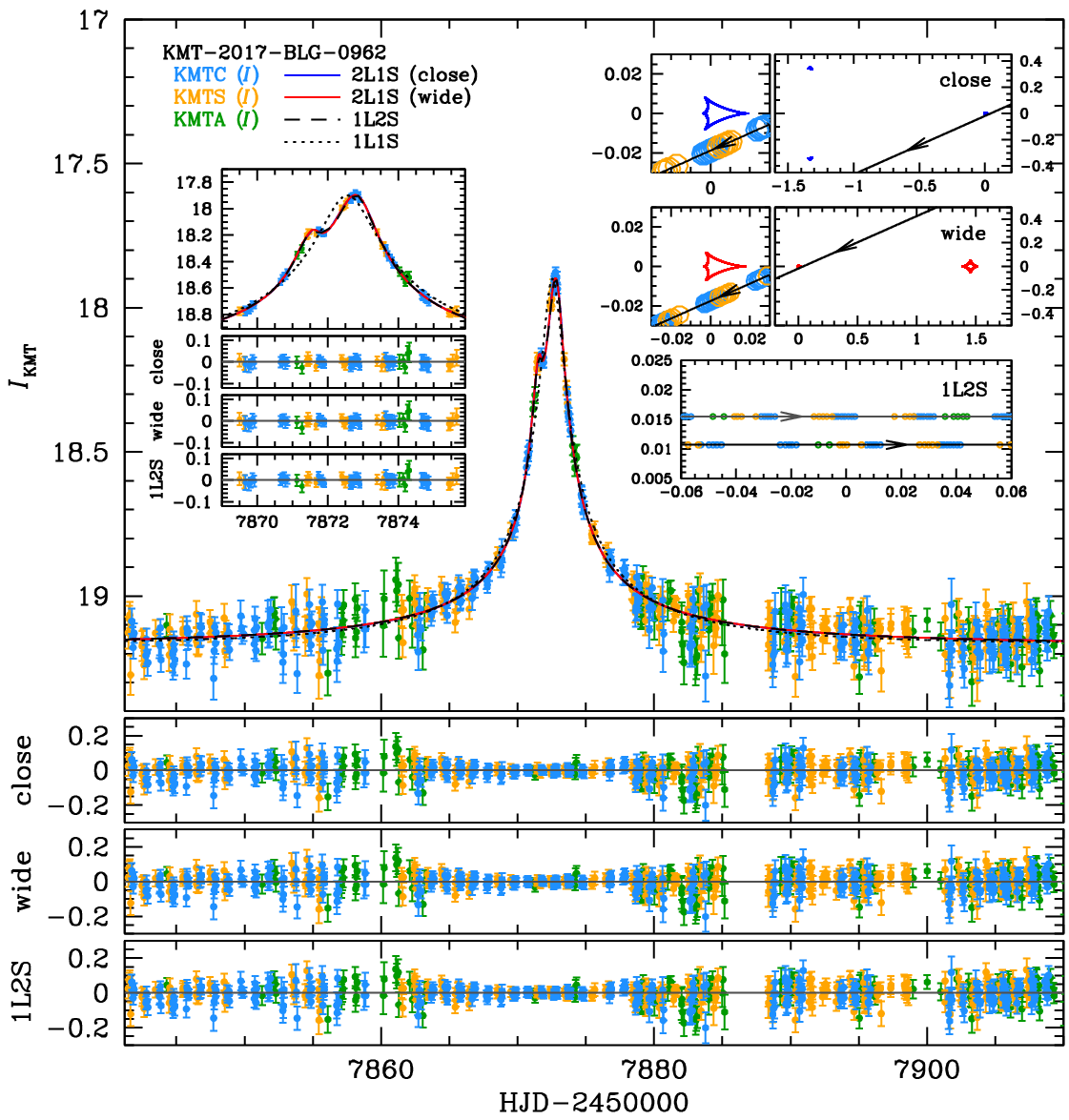}
\caption{
Degenerate models of KMT-2017-BLG-0962. The solid lines in red and blue indicate the 2L1S model light curves of the close and wide cases, respectively. The dashed line in black indicates the 1L2S model light curve. The dotted line indicates the 1L1S model light curve for this event. Left-side inner panels show a zoom-in for the anomaly part of the light curve with residuals. Right-side inner panels present geometries of 2L1S (upper and middle panels for the close and wide cases) and 1L2S (bottom panel) models. Three bottom panels show residuals between each model and observations. 
\label{fig:lc0962}}
\end{figure*}

 However, we found that KMT-2017-BLG-1119 was located within the footprint of another survey. The Microlensing Observations in Astrophysics \citep[MOA:][]{sumi03} survey observed this event using the $1.8$ m MOA-II telescope located at the Mount John Observatory in New Zealand, with the customized filter called MOA-Red filter (wide $R+I$ filter). Because the MOA survey did not alert this event during the $2017$ season, we separately requested the MOA data of the event. The data were reduced using their pipeline employed the difference image analysis (DIA) photometry \citep{bond01}. In contrast, KMT-2017-BLG-0962 is not located in the MOA observation fields.

\begin{figure*}[htb!]
\epsscale{1.00}
\plotone{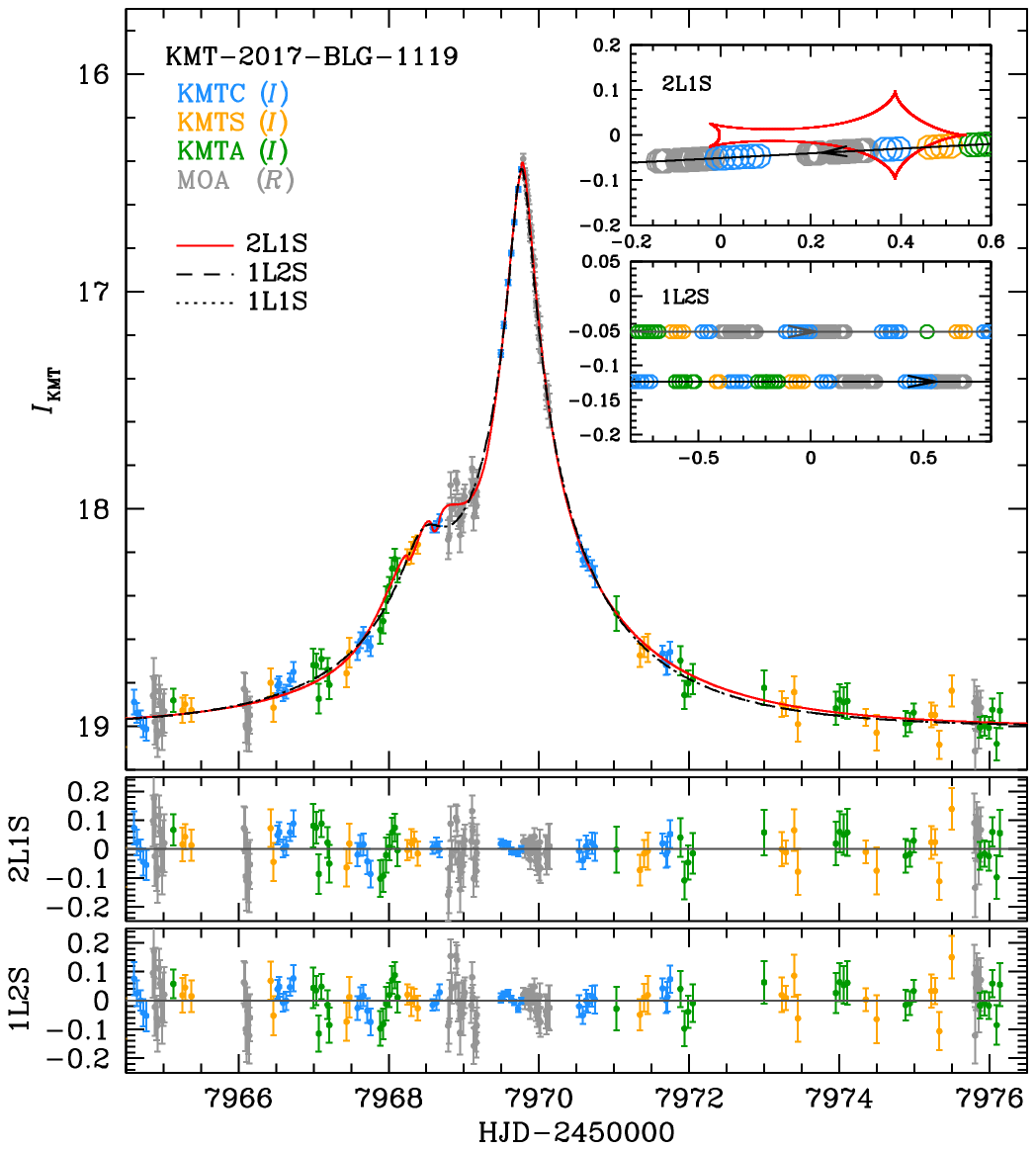}
\caption{
Degenerate models of KMT-2017-BLG-1119. The description is the same as for Figure \ref{fig:lc0962}.
\label{fig:lc1119}}
\end{figure*}

\subsection{KMT-2017-BLG-0962}
 KMT-2017-BLG-0962 occurred on source(s) located at $(\alpha,\delta)_{\rm J2000}=(17^{h}46^{m}48^{s}.54,-26^{\circ}10^{'}48^{''}.07)$ corresponding to the Galactic coordinates $(l,b)=(2.^{\circ}49,1.^{\circ}21)$. This event is located in the KMT-field, BLG18 (see Figure 12 of \citealt{kim18}), which has the nominal observational cadence $1\,{\rm hr}^{-1}$. During the event, the cadence was $1\, {\rm hr}^{-1}$ at KMTC. For the other observations, the cadence was $0.75\, {\rm hr}^{-1}$. In Figure \ref{fig:lc0962}, we present KMTNet observations of this event with a 1L1S model curve as a reference to clearly show the anomaly in the light curve. There exist clear perturbations around the peak of the event, ${\rm HJD}' (={\rm HJD}-2450000)\sim 7871.5$.

\subsection{KMT-2017-BLG-1119}
 KMT-2017-BLG-1119 occurred on source(s) located at $(\alpha,\delta)_{\rm J2000}=(17^{h}52^{m}10^{s}.63,-33^{\circ}01^{'}05^{''}.30)$ corresponding to the Galactic coordinates $(l,b)=(-2.^{\circ}78,-3.^{\circ}30)$. This event is located in the KMT-field, BLG22, which also has the nominal cadence $1\,{\rm hr}^{-1}$. During the event, this was an actual cadence for KMTC observations. For KMTS and KMTA observations, the cadence was switched from $1\,{\rm hr}^{-1}$ to $0.75\,{\rm hr}^{-1}$ at ${\rm HJD'}\sim7971.25$, i.e., just after the event peaked. In Figure \ref{fig:lc1119}, we present the KMTNet and MOA observations of this event. The observations show clear deviations (from ${\rm HJD'}\sim 7967.5$ to $\sim 7969.0$) from the 1L1S model.

\section{Interpretations of the Light curves}
 Because both events show clear anomalies in the observed light curves, we analyze the light curves using both 2L1S and 1L2S interpretations. For each interpretation, we build model light curves using an appropriate parameterization. Then we minimize the $\chi^2$ difference between the model and observations by using a Markov Chain Monte Carlo (MCMC) algorithm \citep{dunkley05}.

\begin{deluxetable*}{lrrlr}
\tablecaption{Best-fit parameters of degenerate models of KMT-2017-BLG-0962\label{table:0962}}
\tablewidth{0pt}
\tablehead{
\multicolumn{1}{c}{parameter} &
\multicolumn{1}{c}{2L1S (close)} &
\multicolumn{1}{c}{2L1S (wide)} &
\multicolumn{1}{c}{parameter} &
\multicolumn{1}{c}{1L2S} 
}
\startdata
$\chi^2 / {\rm N_{data}}$  & $1918.423/1918$              & $1918.687/1918$              & $\chi^2 / {\rm N_{data}}$ & $1919.026/1918$              \\
$t_0$ (HJD')               & $7872.514_{-0.015}^{+0.009}$ & $7872.536_{-0.011}^{+0.011}$ & $t_{0,S1}$                & $7871.478_{-0.029}^{+0.024}$ \\
$u_0$                      & $   0.017_{-0.002}^{+0.003}$ & $   0.016_{-0.001}^{+0.002}$ & $t_{0,S2}$                & $7872.797_{-0.024}^{+0.012}$ \\
$t_{\rm E}$ (days)         & $  33.380_{-4.002}^{+2.966}$ & $  35.513_{-4.445}^{+3.026}$ & $t_{\rm E}$               & $  34.435_{-4.044}^{+3.215}$ \\
$s$                        & $   0.529_{-0.048}^{+0.012}$ & $   1.964_{-0.069}^{+0.210}$ & $u_{0,S1}$                & $   0.011_{-0.002}^{+0.002}$ \\
$q$                        & $   0.012_{-0.002}^{+0.004}$ & $   0.011_{-0.002}^{+0.004}$ & $u_{0,S2}$                & $   0.015_{-0.001}^{+0.002}$ \\
$\alpha$                   & $   2.723_{-0.011}^{+0.006}$ & $   2.725_{-0.011}^{+0.006}$ & $q_{\rm flux}$            & $   4.099_{-0.489}^{+0.844}$ \\
$\rho_{\ast}$              & $   \leq 0.010             $ & $   \leq 0.009             $ & $\rho_{\ast}$             & \nodata                      \\
$F_{\rm S,KMTC}$           & $   0.014_{-0.001}^{+0.002}$ & $   0.013_{-0.001}^{+0.002}$ & $F_{\rm S,KMTC}$          & $   0.014_{-0.001}^{+0.002}$ \\
$F_{\rm B,KMTC}$           & $   0.327_{-0.002}^{+0.001}$ & $   0.328_{-0.002}^{+0.001}$ & $F_{\rm B,KMTC}$          & $   0.327_{-0.002}^{+0.001}$ \\
$F_{\rm S,KMTS}$           & $   0.013_{-0.001}^{+0.002}$ & $   0.012_{-0.001}^{+0.002}$ & $F_{\rm S,KMTC}$          & $   0.013_{-0.001}^{+0.002}$ \\
$F_{\rm B,KMTS}$           & $   0.329_{-0.002}^{+0.001}$ & $   0.330_{-0.002}^{+0.001}$ & $F_{\rm B,KMTC}$          & $   0.329_{-0.002}^{+0.001}$ \\
$F_{\rm S,KMTA}$           & $   0.011_{-0.001}^{+0.002}$ & $   0.010_{-0.001}^{+0.002}$ & $F_{\rm S,KMTC}$          & $   0.010_{-0.001}^{+0.001}$ \\
$F_{\rm B,KMTA}$           & $   0.312_{-0.002}^{+0.001}$ & $   0.313_{-0.002}^{+0.001}$ & $F_{\rm B,KMTC}$          & $   0.313_{-0.001}^{+0.001}$ \\
\enddata
\tablecomments{
We present upper limits ($3\sigma$) of the $\rho_{\ast}$ parameters for the 2L1S models. 
Because this event does not have caustic-crossings, the $\rho_{\ast}$ parameters are not 
accurately measured (see Figure \ref{fig:dist0962}). For the 1L2S models, the finite 
source effect is not considered for modeling.
}
\end{deluxetable*}

 During the modeling process, the uncertainties of observations are rescaled using the equation, $e_{\rm new} = {\kappa}_{\rm obs}\,e_{\rm old}$, where the $e_{\rm new}$ and $e_{\rm old}$ are rescaled and original uncertainties in magnitudes, respectively.\footnotetext[3]{In general, the error rescaling is used a quadrature formalism: $e_{\rm new} = \kappa \sqrt {e_{\rm old}^2 + e_{\rm min}^2 }$, where $\kappa$ and $e_{\rm min}$ are error rescaling factors. However, we find that the $e_{\rm min}$ factors should be zero for observations of both events. Thus, we only present the $\kappa$ factor without the meaningless zero terms.\\ $~~~~~$ In principle, the $\kappa$ factor has an uncertainty of $(2N)^{-1/2}$. Neglecting this factor can affect the interpretation of the $\Delta\chi^2$ difference between two models. Specifically, it leads to an uncertainty in the $\Delta\chi^2$ of $\sqrt{(2/N)}$. Hence, $\sigma(\Delta\chi^2)/\Delta\chi^2 = \sqrt{(2/N)} \rightarrow 3.7\%$ for $N \sim 1500$, so that for example $\Delta\chi^2=10$ would formally be written $\Delta\chi^2=10\pm 0.37$ (for three observatories). This uncertainty in $\Delta\chi^2$ has no practical impact in the present case, so we suppress it in all expressions.} The coefficient $\kappa_{\rm obs}$, an error rescaling factor for each dataset, is defined based on the best-fit model having the lowest $\chi^2$ value. By making sure each data point contributes on average $\Delta\chi^{2}\sim1$, we can quantitatively compare the degenerate models. For KMT-2017-BLG-0962 and KMT-2017-BLG-1119, the sets of error rescaling factors are $(\kappa_{\rm KMTC}, \kappa_{\rm KMTS}, \kappa_{\rm KMTA})=(1.244, 1.239, 1.392)$ and $(\kappa_{\rm KMTC}, \kappa_{\rm KMTS}, \kappa_{\rm KMTA}, \kappa_{\rm MOA})=(1.2208, 1.1209, 1.2017, 0.8930)$, respectively.

\subsection{Parameterization of the 2L1S Interpretation}
 To build a standard 2L1S model light curve, seven basic parameters are required to describe the caustic form and the source trajectory. Two parameters ($s$ and $q$) determine the caustic form. The value $s$ represents the projected separation between the lenses in units of the angular Einstein ring radius ($\theta_{\rm E}$). Conventionally, cases of $s<1$ and $s>1$ are referred to ``close'' and ``wide'', respectively. The mass ratio of the lenses is defined as $q=M_{2}/M_{1}$ where $M_{1}$ and $M_{2}$ are masses of first and second bodies, respectively. These close and wide cases can yield a close/wide degeneracy caused by similarities in the magnification pattern, which are induced by an intrinsic symmetry in the lens equation \citep{griest98,dominik99}.   

 Four parameters ($t_0$, $u_0$, $t_{\rm E}$, and $\alpha$) describe the source trajectory: $t_0$ is the time when the source most closely approaches to the reference position of the lens system (this reference position is the photo-center \citep{kim09} defined as $s[1-(1+q)^{-1}]$ and $s^{-1}q/(1+q)$ for the close ($s<1$) and wide ($s>1$) cases, respectively), $u_0$ is the separation at the time of $t_0$, $t_{\rm E}$ is the Einstein timescale defined as the time for the source to cross the Einstein ring radius of the event, and $\alpha$ is the angle of the source trajectory with respect to the binary axis of the lens system. The geometry of a microlensing event produced by 2L1S is built using these six parameters, which determine the magnification as a function of time, i.e., the microlensing light curve. The finite angular size of the source moderates the magnification. To account for the finite source effect, the final parameter, $\rho_{\ast}$, is required, which is defined as the angular source radius ($\theta_{\ast}$) scaled by $\theta_{\rm E}$. In addition, we introduce two additional parameters, $F_{\rm S, obs}$ (source flux) and $F_{\rm B, obs}$ (blending flux), for each dataset, which are used to scale the model to the data. These parameters are determined based on the model using the least-square fitting method.

\subsection{Parameterization of the 1L2S Interpretation}
 A standard 1L2S model light curve is built using a superposition of two 1L1S light curves induced by each source. The trajectory of each source yields the individual magnification of its 1L1S light curve. For the 1L2S model light curve, the final magnification is calculated by superposing magnifications of both sources weighted by the flux ratio of source stars. To describe the source trajectories, there are two parameterizations. The first parameterization (hereafter, A-type) describes the trajectory of each source, individually. In contrast, the second parameterization (hereafter, B-type) describes the barycenter motion of the binary-source system. Then, from the position of the barycenter, the position of each source is derived. In Appendix A, we provide detailed descriptions of these parameterizations and discuss the pros and cons of the two types. In this work, because the merits of the two types are different, we adopt the A-type for the basic 1L2S modeling (Section 3.3 and 3.4). For testing the higher-order effects, we adopt the B-type (Section 3.4).

\subsection{Degenerate Models}

\begin{deluxetable*}{lrlr}
\tablecaption{Best-fit parameters of degenerate models of KMT-2017-BLG-1119 \label{table:1119}}
\tablewidth{0pt}
\tablehead{
\multicolumn{1}{c}{parameter} &
\multicolumn{1}{c}{2L1S} &
\multicolumn{1}{c}{parameter} &
\multicolumn{1}{c}{1L2S} 
}
\startdata
$\chi^2 / {\rm N_{data}}$  & $1580.372/1579$              & $\chi^2 / {\rm N_{data}}$ & $1610.081/1579$              \\
$t_0$ (HJD')               & $7969.731_{-0.006}^{+0.002}$ & $t_{0,S1}$                & $7968.468_{-0.034}^{+0.044}$ \\
$u_0$                      & $   0.051_{-0.001}^{+0.003}$ & $t_{0,S2}$                & $7969.769_{-0.002}^{+0.003}$ \\
$t_{\rm E}$ (days)         & $   2.917_{-0.110}^{+0.048}$ & $t_{\rm E}$               & $   2.449_{-0.189}^{+0.137}$ \\
$s$                        & $   1.211_{-0.001}^{+0.016}$ & $u_{0,S1}$                & $  -0.123_{-0.032}^{+0.020}$ \\
$q$                        & $   0.009_{-0.001}^{+0.001}$ & $u_{0,S2}$                & $  -0.051_{-0.005}^{+0.003}$ \\
$\alpha$                   & $   3.089_{-0.006}^{+0.001}$ & $q_{\rm flux}$            & $   4.720_{-0.679}^{+0.526}$ \\
$\rho_{\ast}$              & $   0.029_{-0.001}^{+0.003}$ & $\rho_{\ast}$             & \nodata                      \\
$F_{\rm S,KMTC}$           & $   0.185_{-0.006}^{+0.007}$ & $F_{\rm S,KMTC}$          & $   0.248_{-0.019}^{+0.030}$ \\
$F_{\rm B,KMTC}$           & $   0.210_{-0.007}^{+0.006}$ & $F_{\rm B,KMTC}$          & $   0.147_{-0.032}^{+0.017}$ \\
$F_{\rm S,KMTS}$           & $   0.157_{-0.012}^{+0.003}$ & $F_{\rm S,KMTC}$          & $   0.206_{-0.023}^{+0.033}$ \\
$F_{\rm B,KMTS}$           & $   0.256_{-0.004}^{+0.012}$ & $F_{\rm B,KMTC}$          & $   0.208_{-0.034}^{+0.021}$ \\
$F_{\rm S,KMTA}$           & $   0.145_{-0.009}^{+0.007}$ & $F_{\rm S,KMTC}$          & $   0.212_{-0.020}^{+0.031}$ \\
$F_{\rm B,KMTA}$           & $   0.251_{-0.007}^{+0.009}$ & $F_{\rm B,KMTC}$          & $   0.184_{-0.032}^{+0.018}$ \\
$F_{\rm S,MOA}$            & $   0.158_{-0.002}^{+0.012}$ & $F_{\rm S,MOA}$           & $   0.221_{-0.017}^{+0.024}$ \\
$F_{\rm B,MOA}$            & $   0.242_{-0.012}^{+0.002}$ & $F_{\rm B,MOA}$           & $   0.179_{-0.024}^{+0.014}$ \\
\enddata
\tablecomments{
For the 1L2S models, the finite source effect is not considered for modeling.
}
\end{deluxetable*}

\subsubsection{KMT-2017-BLG-0962}
 For KMT-2017-BLG-0962, we find that the observed light curve can be described using either 2L1S and 1L2S interpretations. In Figure \ref{fig:lc0962}, we present the observed data and model light curves of this event with geometries of the 2L1S and 1L2S interpretations. We also present residuals between the models and observations. In Table \ref{table:0962}, we present the model parameters of best-fit models with $\chi^2$ between the models and observations. The 2L1S model indicates that this event can be caused by a planetary lens system with a mass ratio $q\sim 0.01$ between the lens components. However, there is a degeneracy between the close and wide solutions. At the same time, the 1L2S model implies that the event can also be caused by a binary-source system. The planetary model (2L1S models of the close and wide cases) and 1L2S model are completely degenerate. The $\chi^2$ differences between 1L2S and 2L1S are only $\sim0.6$ and $\sim0.3$ for the close and wide cases, respectively. Thus, we cannot claim a certain planet discovery.

\subsubsection{KMT-2017-BLG-1119}
 For KMT-2017-BLG-1119, we find that the observed light curve is also well-described by both interpretations. In Figure \ref{fig:lc1119}, we present light curves of these degenerate models with their geometries and residuals. In Table \ref{table:1119}, we present the parameters of these degenerate models. In contrast to the previous case, these models show slight variations. The best-fit model, 2L1S, shows a low mass ratio ($q\sim0.009$) with very short Einstein timescale ($t_{\rm E}\sim 2.92$ days). This indicates that this event can be caused by a low-mass planetary lens system. However, this event also can be well described by the 1L2S interpretation, which implies that the planet would not exist. Quantitatively, the $\chi^2$ difference between 1L2S and 2L1S is $\Delta\chi^{2}\sim29.7$. This $\Delta\chi^2$ value is too marginal to claim the 2L1S/2L1S degeneracy is resolved considering the severe systematics of the observations (see residuals of Figure 2). The $\Delta\chi^2$ cannot be conclusive evidence to resolve the degeneracy (we discuss more details of the $\chi^2$ difference in Section 4.1).

\subsection{Higher-order Effects of the Interpretations}
 Even though both events have the 2L1S/1L2S degeneracy, it is possible that these events were caused by planetary lens systems. Thus, for the 2L1S interpretation, we check the possibility of measuring the annual microlens parallax \citep[APRX:][]{gould92} because the microlens parallax is not only a key observable for directly determining the properties of the lens system but also a strong constraint for estimating the properties using the Bayesian analysis. However, we cannot find any meaningful improvements for both events to claim the detection of the APRX signals when we consider the APRX models by introducing the additional parameters of the microlens parallax. 

 For the 1L2S interpretation, the binary sources orbit each other and conserve their angular momentum. This source-orbital motion can affect the light curve. In addition, the source-orbital effect can be a clue to resolving the 2L1S/1L2S degeneracy. Thus, we test the effect of the source-orbital motion by adopting the B-type parameterization with two additional orbital parameters (see Appendix A for details of this parameterization). However, we cannot find any meaningful signals in the light curves of both events caused by the orbital motion of the sources (for the details of non-detection of these higher-order effects, see Appendix B).

\section{Resolving the Degeneracy}

\subsection{Detailed Analysis of the Light curve}
 We now consider whether the 2L1S/1L2S degeneracy can be resolved in either of the two events. There are several methods that may be employed to resolve this degeneracy, most of which were discussed by \citet{gaudi98}. The first method is the detailed analysis of the light curve to check for small differences between the two models. 

 For KMT-2017-BLG-0962, the $\chi^{2}$ difference between the 2L1S and 1L2S models is insignificant, and Figure \ref{fig:lc0962} shows that the three models are quite similar. In contrast to the \citet{gaudi98} case, there are no caustic-crossings. There exists only a smooth deviation from a 1L1S event. Thus, for this event, the differences in the light curve are not sufficient to resolve the degeneracy.

\begin{figure}[htb!]
\epsscale{1.00}
\plotone{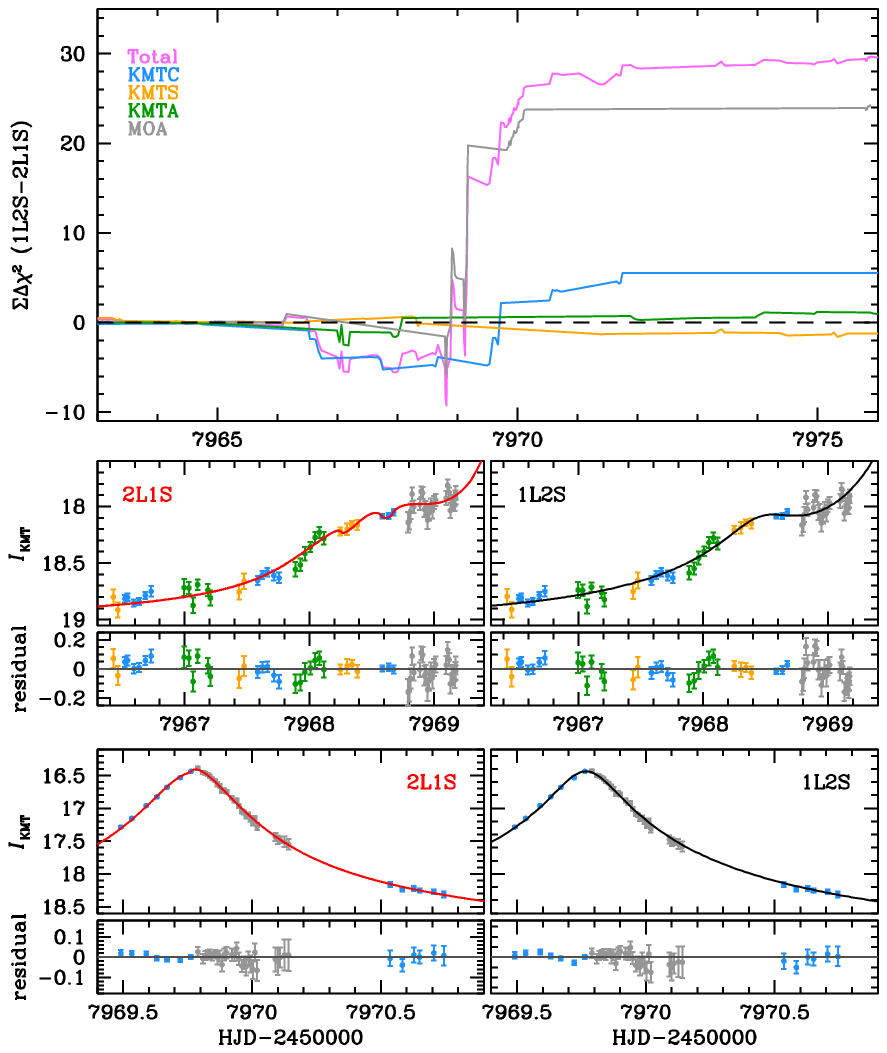}
\caption{
Cumulative $\chi^2$ difference ($\Sigma\Delta\chi^2$) of degenerate models with zoom-ins for anomaly 
part of KMT-2017-BLG-1119. Upper panel shows the $\Sigma\Delta\chi^2$ of total and each dataset.  
Lower four panels present zoom-ins of anomaly parts with residuals of each model case.
\label{fig:dchi2}}
\end{figure}

 For KMT-2017-BLG-1119, the best 2L1S model is preferred by $\Delta\chi^2\sim 30$ over the 1L2S model. However, even though the degeneracy is formally broken, the distinction is not as strong as it appears. In Figure \ref{fig:dchi2}, we present plots of the cumulative $\chi^2$ of each model to investigate the origin of the $\chi^2$ improvement. We find that the $\chi^2$ improvement starts at ${\rm HJD'}\sim7969.0$, which is a part of the light curve covered by MOA and KMTC observations. The $\chi^2$ improvement mostly comes from the MOA observations. Quantitatively, among the total $\chi^2$ improvement, the MOA and KMTC data contribute $\Delta\chi^2 \sim 24$ and $\sim 6$, respectively. However, both datasets have systematics that persist even in the best model (see lower four panels of zoom-in in Figure \ref{fig:dchi2}). This fact suggests that a significant portion of the improvement could just be from fitting systematics in the data. Thus, $\Delta\chi^2$ cannot be a conclusive clue to resolve the 2L1S/1L2S degeneracy. In addition, while this still indicates a preference for the 2L1S model, the physical parameters derived from the Bayesian analysis in Section 5.2.2 predict an extreme system in which the host itself is a massive planet. Thus, we should consider other means of testing the models to independently resolve the degeneracy.

\subsection{Color Information of the Source(s)}

\begin{figure}[htb!]
\epsscale{1.00}
\plotone{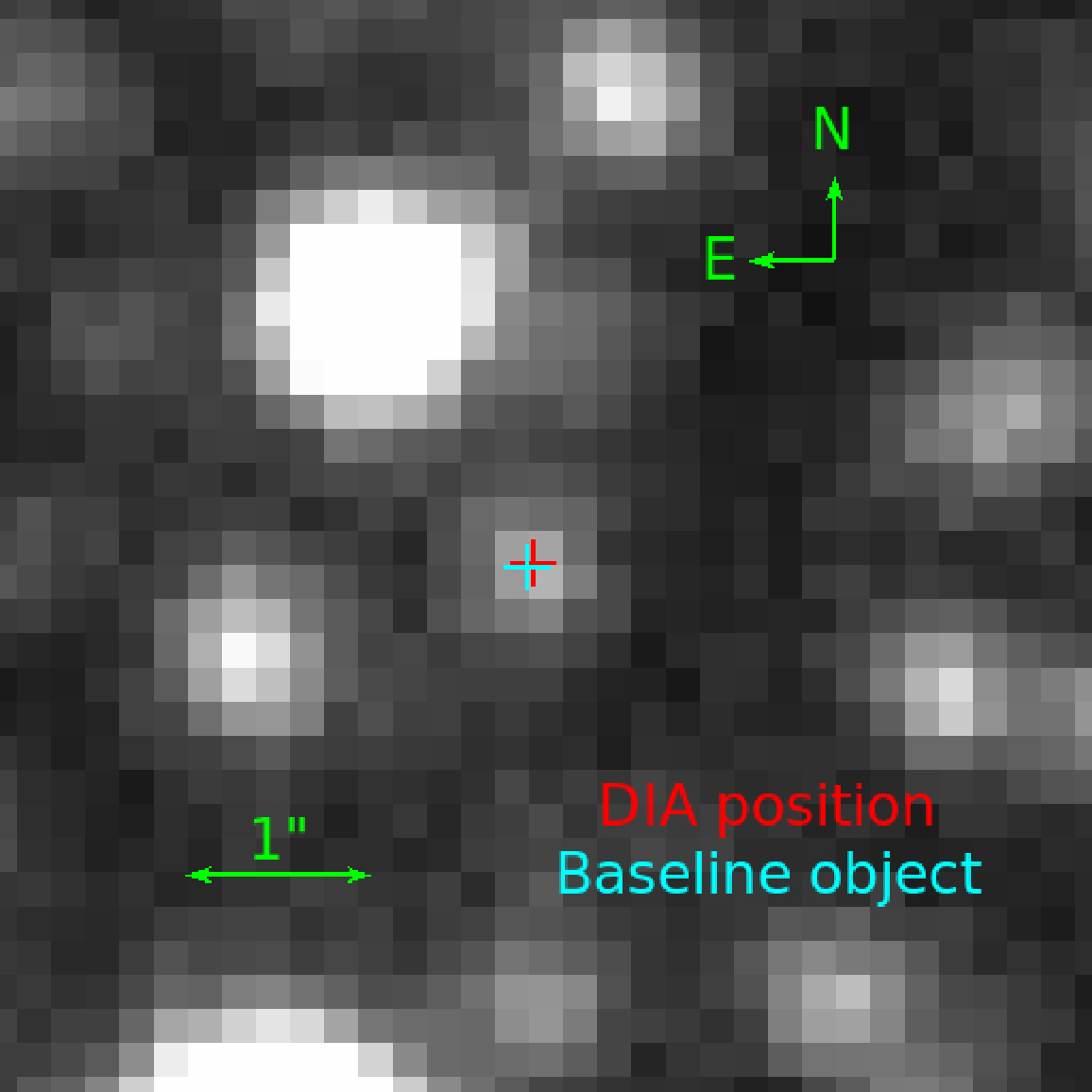}
\caption{
The CFHT image with the astrometric offset ($0.037\pm0.009\arcsec$) between the baseline object positions obtained from the CFHT image (cyan) and the KMTNet catalog (red) that is measured using difference image analysis (DIA). The green arrows indicate the north and east directions (upper-right) and a scale of $\sim1$ arcsecond (lower-left).
\label{fig:offset}}
\end{figure}

 The second method is to use the source-color information. Because the magnification of the 1L2S model is a weighted mean using the flux ratio of the sources (see Appendix A), the final magnification is wavelength dependent. Thus, if the binary sources have different colors (and the event really is a 1L2S event), we can measure the color change or difference during the perturbation from multi-band observations. However, unfortunately, the signal-to-noise ratio of {\it V}-band observations (the KMTNet regularly takes {\it V}-band images) for both events is too low to apply this method. Thus, we cannot resolve the degeneracy using this method.

\subsection{Other Methods to Resolve the Degeneracy}

 \citet{gaudi98} also proposed additional observations to resolve the degeneracy if the previous methods fail. One spectroscopic method requires taking spectra of the source both during and after the perturbations of the event. However, this method cannot be used after the events have ended. The other method requires photometrically and spectroscopically monitoring of the source after the event to search for other signals induced by the binary source such as radial velocity variations due to orbital motion or eclipses. Given the faintness of the source(s), spectroscopic monitoring would be challenging. And given the source separations ($0.04\,\theta_{\rm E}$ and $0.5\,\theta_{\rm E}$ for KMT-2017-BLG-0962 and KMT-2017-BLG-1119, respectively), the probability of eclipses is extremely low. In addition, \citet{calchi18} presented a new method to resolve the 2L1S/1L2S degeneracy using simultaneous ground- and space-based observations. However, unfortunately, space-based data do not exist for these events (see footnote 2).

\subsection{Measurement of the Baseline Object}

 Because most possibilities, which are proposed by other studies, are not helpful to resolve the 2L1S/1L2S degeneracy of our cases, we consider another possibility to resolve the degeneracy using higher-resolution follow-up observations to directly measure the magnitude of the source(s) for these events. For KMT-2017-BLG-1119, we found different source fluxes ($F_{\rm S, KMTC}$) for the 2L1S and 1L2S interpretations (see Table \ref{table:1119}). If this event was caused by the planetary system, the magnitude of the source will be $I=19.85\pm0.04$ and the lens is predicted to be dark. If this event was caused by binary sources, the integrated magnitude of the sources will be observed $I=19.54\pm0.11$. We note that these expected {\it I} magnitudes are calibrated to the OGLE-III magnitude system by cross-matching between KMTNet and OGLE-III catalogs $(I_{\rm OGLE} = (0.0228 \pm 0.0125) + I_{\rm KMTNet})$.

\begin{figure}[htb!]
\epsscale{1.00}
\plotone{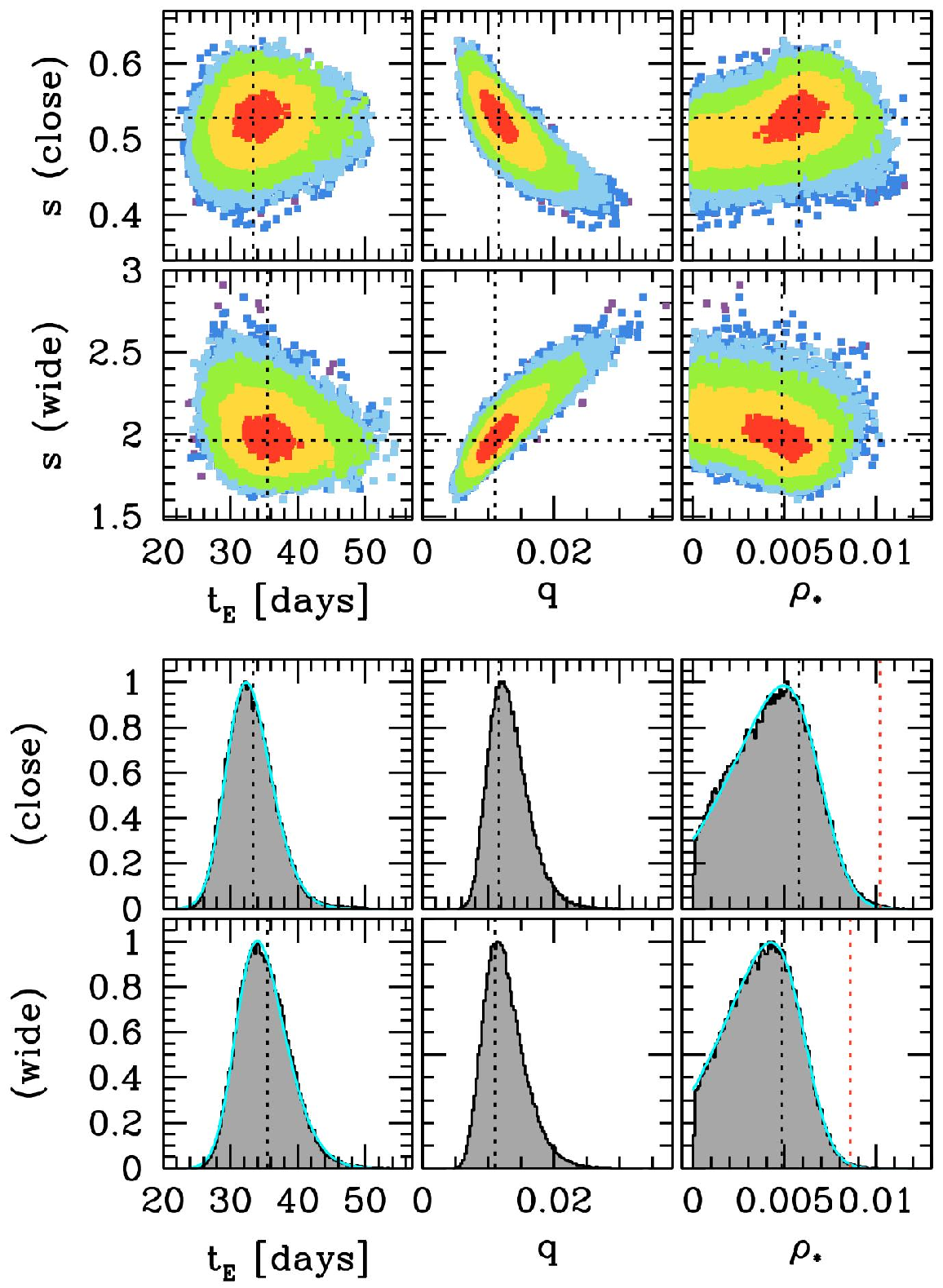}
\caption{
Distributions of $t_{\rm E}$, $q$, and $\rho_{\ast}$ parameters for KMT-2017-BLG-0962. The upper six panels present 2D distributions for the close and wide cases of the 2L1S model obtained from the MCMC chains. Each color represents $\Delta\chi^{2}$ between realization on the chain and the best-fit model: $1^{2}$ (red), $2^{2}$ (yellow), $3^{2}$ (green), $4^{2}$ (sky blue), $5^{2}$ (blue), and $6^{2}$ (purple). The lower six panels present 1D distributions of $t_{\rm E}$, $q$, and $\rho_{\ast}$ parameters for the close and wide cases. The cyan lines indicate weight functions constructed by the fitting of the skewed Gaussian function. The black dotted line indicates the parameter value of the best-fit model. The red dotted line in the $\rho_{\ast}$ distributions represent the $3\sigma$ values.
\label{fig:dist0962}}
\end{figure}

 We check the expected brightness of the baseline object using observations taken from the Canada-France-Hawaii Telescope (CFHT) located at the Maunakea Observatories in $2018$. In Figure \ref{fig:offset}, we present the CFHT image with the astrometric offset between the positions of the baseline object obtained from CFHT and KMTNet observations. The offset is $0.037\arcsec\pm0.009\arcsec$. From the CFHT image, we measure the brightness of the baseline object. We also see that the baseline object is close to coincident with the event and isolated. Thus, it is highly likely that the light from the baseline object is composed of light from stars related to the event. Thus, the CFHT measurement can be a constraint to check the degenerate solutions of this event. From the stacked deep CFHT image (seeing$\sim0.7"$), we can measure the brightness of the baseline object: $I_{\rm base}=19.62\pm0.05$ (we note that the CFHT instrumental magnitude is also calibrated to the OGLE-III magnitude system). The measurement of the baseline object is consistent with the expectation of the 1L2S interpretation considering its $1\sigma$ uncertainty. Therefore, this constraint supports the conclusion that this event might be caused by the 1L2S system. However, we cannot guarantee that the CFHT measurement completely excludes blend light from unrelated stars. Thus, the possibility of the 2L1S origin cannot be clearly ruled out although it is disfavored.

\begin{figure}[htb!]
\epsscale{1.00}
\plotone{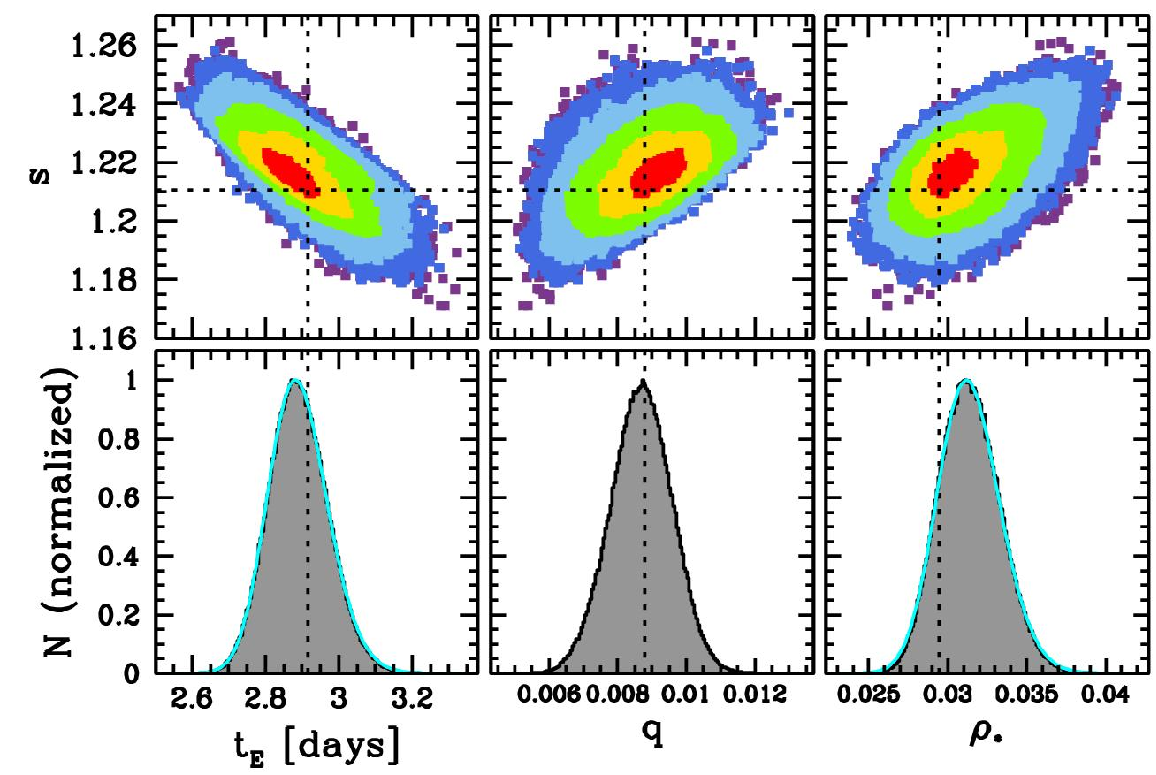}
\caption{
Distributions of $t_{\rm E}$, $q$, and $\rho_{\ast}$ parameters for KMT-2017-BLG-1119. The description is the same as for Figure \ref{fig:dist0962}.
\label{fig:dist1119}}
\end{figure}

 The 2L1S solution predicts a lens-source relative proper motion of $4.7\pm0.6\,{\rm mas\,yr^{-1}}$. Thus, if a thirty-meter class telescope made observations a decade after the event and the relative proper motion of the source and lens were measured to be significantly different from 2L1S value, that would rule out that solution. On the other hand, if the proper motion were consistent with the 2L1S value, that would tend to support the planetary solution but would not be definitive. Note that such a measurement (as always) requires that the lens (or a companion to the lens) be luminous. However, the short timescale of this event favors low-mass lenses, which might fail this condition.

\begin{deluxetable}{crrrrcc}
\tablecaption{The best-fit parameters of weight functions \label{table:weight}}
\tablewidth{0pt}
\tablehead{
\multicolumn{1}{c}{event} &
\multicolumn{4}{c}{KMT-2017-BLG-0962} & 
\multicolumn{2}{c}{KMT-2017-BLG-1119} \\
\multicolumn{1}{c}{model} &
\multicolumn{2}{c}{close} &
\multicolumn{2}{c}{wide}  &
\multicolumn{2}{c}{resonant} \\
\multicolumn{1}{c}{parameter}        &
\multicolumn{1}{r}{$W(t_{\rm E})$}   &
\multicolumn{1}{r}{$W(\rho_{\ast})$} &
\multicolumn{1}{r}{$W(t_{\rm E})$}   &
\multicolumn{1}{r}{$W(\rho_{\ast})$} &
\multicolumn{1}{c}{$W(t_{\rm E})$}   &
\multicolumn{1}{c}{$W(\rho_{\ast})$} 
}
\startdata
$\eta$   &  0.706 &  0.593 &  0.672 &  0.597 & 0.766 & 0.780 \\
$\mu$    & 29.780 &  0.007 & 31.103 &  0.006 & 2.826 & 0.030 \\
$\sigma$ &  4.745 &  0.004 &  5.452 &  0.004 & 0.105 & 0.003 \\
$\alpha$ &  1.683 & -3.077 &  2.022 & -3.164 & 1.285 & 1.201 \\
\enddata
\end{deluxetable}

 In contrast, for KMT-2017-BLG-0962, we obtained almost identical values of the $F_{\rm S}$ (see Table \ref{table:0962}). Thus, for this event, the measurement of the baseline object using higher-resolution follow-up observations would not be helpful for resolving the degeneracy.

\section{Properties of Planet Candidates}

\subsection{Bayesian Analyses}
 Because we cannot measure the microlens parallax, we estimate the properties of these planet candidates using the Bayesian analyses. We build a prior by generating artificial microlensing events (the total number of simulated events is $4\times10^{7}$). To generate these events, we adopt the Galactic models from various studies: initial and present-day mass functions of \citet{chabrier03}, velocity distributions of \citet{hangould95}, and matter density profiles of the Galactic bulge and disk of \citet{hangould03}. When these artificial microlensing events are generated, the line of sight to the actual event is considered. This prior contains various information about host properties according to the event rate. Based on the event rate, we calculate the posterior probability distributions of the lens properties, by applying constraints obtained from the actual event.

\begin{figure*}[htb!]
\epsscale{1.00}
\plotone{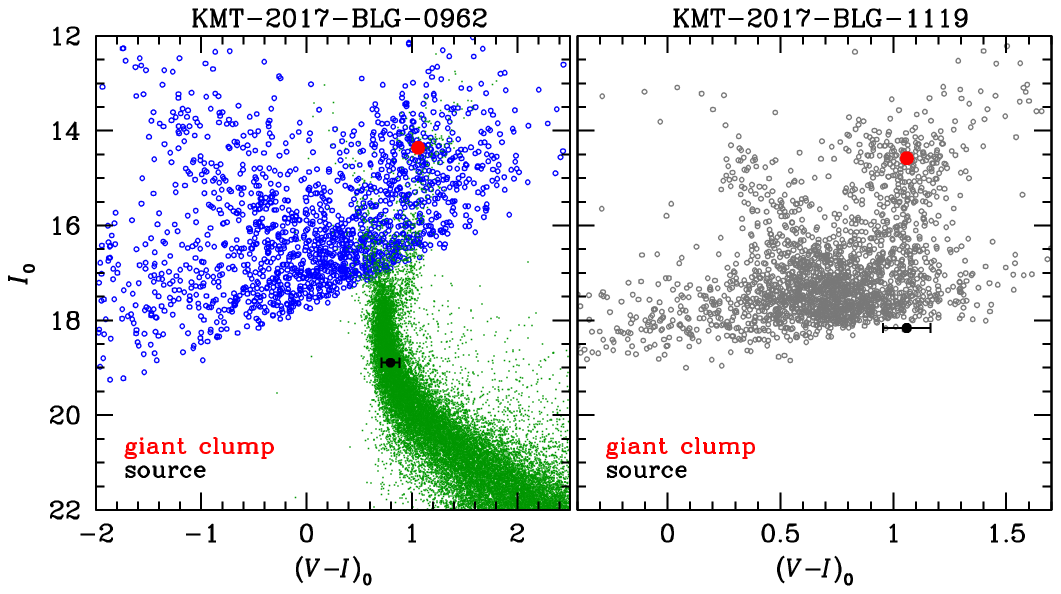}
\caption{
Combined color-magnitude diagrams of KMT-2017-BLG-0962 (left) and KMT-2017-BLG-1119 (right), which are corrected for reddening. The green dots show the CMD of the Galactic bulge observed by the {\it Hubble Space Telescope} \citep{holtzman98}. The blue dots show the CMD of KMTNet constructed using pyDIA reductions. The gray dots show the KMTNet CMD de-reddened and converted to the OGLE-III magnitude system. The red and black dots indicate the centroid of the red giant clump and the estimated source of each event, respectively.
\label{fig:cmd}}
\end{figure*}

 The constraints are built in the form of weight functions, which are obtained from the $t_{\rm E}$ and $\rho_{\ast}$ distributions of the actual event. In Figures \ref{fig:dist0962} and \ref{fig:dist1119}, we present the distributions of selected parameters ($t_{\rm E}$, $\rho_{\ast}$, and $q$), the first two of which are used to build the weight functions and determine the lens properties for KMT-2017-BLG-0962 and KMT-2017-BLG-1119, respectively. The distributions show a skewed Gaussian form, which we parameterize by,
\begin{equation}
W(x)={\eta} e^{{-\frac{1}{2}\left(\frac{x-\mu}{\sigma} \right)^{2}}}\left\{1+{\rm erf} \left[ {\frac{\alpha(x-\mu)}{\sqrt{2}\sigma}} \right] \right\},
\end{equation}
where the function ${\rm erf[z]}$ indicates an error function defined as ${\rm erf[z]}=(1/\sqrt{\pi})\int_{-z}^{z}{e^{-t^{2}}}dt$. The variable $x$ is $t_{\rm E}$ or $\rho_{\ast}$. The set of ($\eta$, $\mu$, $\sigma$, $\alpha$) are fitting parameters. We use the MCMC algorithm to fit these parameters. The fitting results, i.e., $t_{\rm E}$ and $\rho_{\ast}$ weight functions, $W(t_{\rm E})$ and $W(\rho_{\ast})$, are presented in Figures \ref{fig:dist0962} and \ref{fig:dist1119} (cyan lines). In Table \ref{table:weight}, we present the best-fit parameter sets of $W(t_{\rm E})$ and $W(\rho_{\ast})$ for both events. The final weight function is $W=W(t_{\rm E})W(\rho_{\ast})$. By applying the final weight function to the event rate, we construct probability distributions of the host mass ($M_{\rm L}$), the distance to the lens ($D_{\rm L}$), the physical Einstein ring radius ($r_{\rm E}$), and the lens-source relative proper motion ($\mu_{\rm rel}$). From these probability distributions, we can determine the properties of the planet candidate of each event.

\subsection{Angular Source Radius}
 For applying the $W(\rho_{\ast})$ to the event rate, the angular source radius ($\theta_{\ast}$) is required to convert from $\theta_{\rm E}$ (for the artificial lensing events) to $\rho_{\ast}$ ($\rho_{\ast}=\theta_{\ast}/\theta_{\rm E}$). However, unfortunately, we do not have reliable {\it V}-band data to estimate $\theta_{\ast}$. Thus, we cannot adopt the conventional method \citep{yoo04} using ($V-I$) color of the source for measuring the $\theta_{\ast}$. For each event, we estimate $\theta_{\ast}$ using different methods because the available observations are different.

\subsubsection{KMT-2017-BLG-0962}
 For this event, reliable observations to measure the source color do not exist. Thus, we adopt a statistical method \citep[established in][]{bennett08} to estimate the source color using {\it Hubble Space Telescope (HST)} observations of Baade's window \citep{holtzman98}.

 The source magnitude offset from the red giant clump ($\Delta I_{\rm S}= 4.531\pm0.110$) is determined from comparing the source flux ($F_{\rm S,pyDIA}$) obtained from the pyDIA light curve to the red giant clump centroid measured from the color-magnitude diagram (CMD). Then, we extract {\it HST} stars having similar magnitude offset to those of the source of the event. Using this extracted {\it HST} star sample (and excluding $3\sigma$ outliers in $V-I$), we determine the median star color ($<(V-I)_{HST}>$) and the standard deviation of the color ($\sigma(V-I)_{HST}$). Then, we take this {\it HST} star color with uncertainty as representative of the source color: $(V-I)_{\rm S} = 1.357\pm0.083$. By adopting the clump color for the HST CMD from \citet{bennett08}, we find the offset of the source from the clump is $\Delta(V-I) = -0.263\pm0.083$. Then, using the intrinsic color \citep[$1.06;$][]{bensby11} and magnitude \citep[$14.362;$][]{nataf13} of the red giant clump along this line of sight, we derive: $(V-I, I)_{\rm S, 0} = (0.797\pm0.083, 18.893\pm0.110)$. Lastly, $\theta_{\ast}$ is estimated using the color/surface-brightness relation adopted from \citet{kervella04}: 
\begin{equation}
\theta_{\ast} = 0.58\pm0.06 \,{\mu{\rm as}} .
\end{equation}
In Figure \ref{fig:cmd}, we present the combined CMDs of events where the centroids of the red giant clumps are aligned to the unextincted red giant clump magnitudes.

\begin{deluxetable*}{lrrrrc}
\tablecaption{Properties of Planetary System Candidates \label{table:properties}}
\tablewidth{0pt}
\tablehead{
\multicolumn{1}{c}{event} &
\multicolumn{4}{c}{KMT-2017-BLG-0962} & 
\multicolumn{1}{c}{KMT-2017-BLG-1119} \\
\multicolumn{1}{c}{constraints} &
\multicolumn{2}{c}{$t_E + \theta_{\rm E}$ } &
\multicolumn{2}{c}{$t_E$ only } &
\multicolumn{1}{c}{$t_E + \theta_{\rm E}$ } \\ 
\multicolumn{1}{c}{model} &
\multicolumn{1}{c}{close} &
\multicolumn{1}{c}{wide} &
\multicolumn{1}{c}{close} &
\multicolumn{1}{c}{wide} &
\multicolumn{1}{c}{resonant} 
}
\startdata
 w/ stellar remnants           &                        &                        &                        &                        &                           \\
$M_{\rm host}$ $(M_{\odot})$   & $0.46_{-0.29}^{+0.34}$ & $0.48_{-0.30}^{+0.34}$ & $0.50_{-0.31}^{+0.34}$ & $0.52_{-0.31}^{+0.34}$ & $0.017_{-0.011}^{+0.041}$ \\
$M_{\rm planet}$ $(M_{\rm J})$ & $5.6 _{-3.7 }^{+4.7 }$ & $5.6 _{-3.6 }^{+4.5 }$ & $6.1 _{-3.8 }^{+4.7 }$ & $6.0 _{-3.7 }^{+4.6 }$ & $0.16 _{-0.10 }^{+0.38 }$ \\  
$D_{\rm L}$ (kpc)              & $6.4 _{-1.8 }^{+1.3 }$ & $6.4 _{-1.8 }^{+1.3 }$ & $6.2 _{-1.8 }^{+1.3 }$ & $6.2 _{-1.8 }^{+1.3 }$ & $8.2  _{-1.1  }^{+1.1  }$ \\
$a_{\perp}$ (au)               & $1.2 _{-0.5 }^{+0.5 }$ & $4.7 _{-1.9 }^{+1.9 }$ & $1.3 _{-0.5 }^{+0.5 }$ & $5.0 _{-1.9 }^{+1.8 }$ & $0.36 _{-0.06 }^{+0.07 }$ \\
$a_{\rm snow}$ (au)            & $1.2 _{-0.8 }^{+0.9 }$ & $1.3 _{-0.8 }^{+0.9 }$ & $1.4 _{-0.8 }^{+0.9 }$ & $1.4 _{-0.8 }^{+0.9 }$ & $0.05 _{-0.03 }^{+0.11 }$ \\
$\mu$ $({\rm mas~yr^{-1}})$    & $4.3 _{-1.9 }^{+2.4 }$ & $4.2 _{-1.9 }^{+2.4 }$ & $4.8 _{-2.0 }^{+2.4 }$ & $4.7 _{-2.0 }^{+2.4 }$ & $4.7  _{-0.6  }^{+0.6  }$ \\
\hline                                                                                                           
w/o stellar remnants           &                        &                        &                        &                        &                           \\
$M_{\rm host}$ $(M_{\odot})$   & $0.38_{-0.24}^{+0.40}$ & $0.40_{-0.25}^{+0.41}$ & $0.43_{-0.26}^{+0.40}$ & $0.44_{-0.27}^{+0.41}$ &                           \\
$M_{\rm planet}$ $(M_{\rm J})$ & $4.7 _{-3.0 }^{+5.2 }$ & $4.7 _{-2.9 }^{+5.1 }$ & $5.2 _{-3.2 }^{+5.3 }$ & $5.1 _{-3.2 }^{+5.1 }$ &                           \\
$D_{\rm L}$ (kpc)              & $6.4 _{-1.9 }^{+1.3 }$ & $6.3 _{-1.9 }^{+1.3 }$ & $6.2 _{-1.9 }^{+1.3 }$ & $6.1 _{-1.9 }^{+1.4 }$ &                           \\
$a_{\perp}$ (au)               & $1.1 _{-0.5 }^{+0.5 }$ & $4.3 _{-1.7 }^{+2.0 }$ & $1.2 _{-0.5 }^{+0.5 }$ & $4.7 _{-1.8 }^{+2.0 }$ &                           \\
$a_{\rm snow}$ (au)            & $1.0 _{-0.6 }^{+1.1 }$ & $1.1 _{-0.7 }^{+1.1 }$ & $1.2 _{-0.7 }^{+1.1 }$ & $1.2 _{-0.7 }^{+1.1 }$ &                           \\
$\mu$ $({\rm mas~yr^{-1}})$    & $4.1 _{-1.9 }^{+2.5 }$ & $4.1 _{-1.8 }^{+2.4 }$ & $4.6 _{-2.0 }^{+2.4 }$ & $4.5 _{-2.0 }^{+2.4 }$ &                           \\
\enddata
\tablecomments{
For KMT-2017-BLG-1119, the median values with and without stellar remnant hosts are identical. Thus, we present one case to avoid clutter.
}
\end{deluxetable*}

\begin{figure}[htb!]
\epsscale{1.10}
\plotone{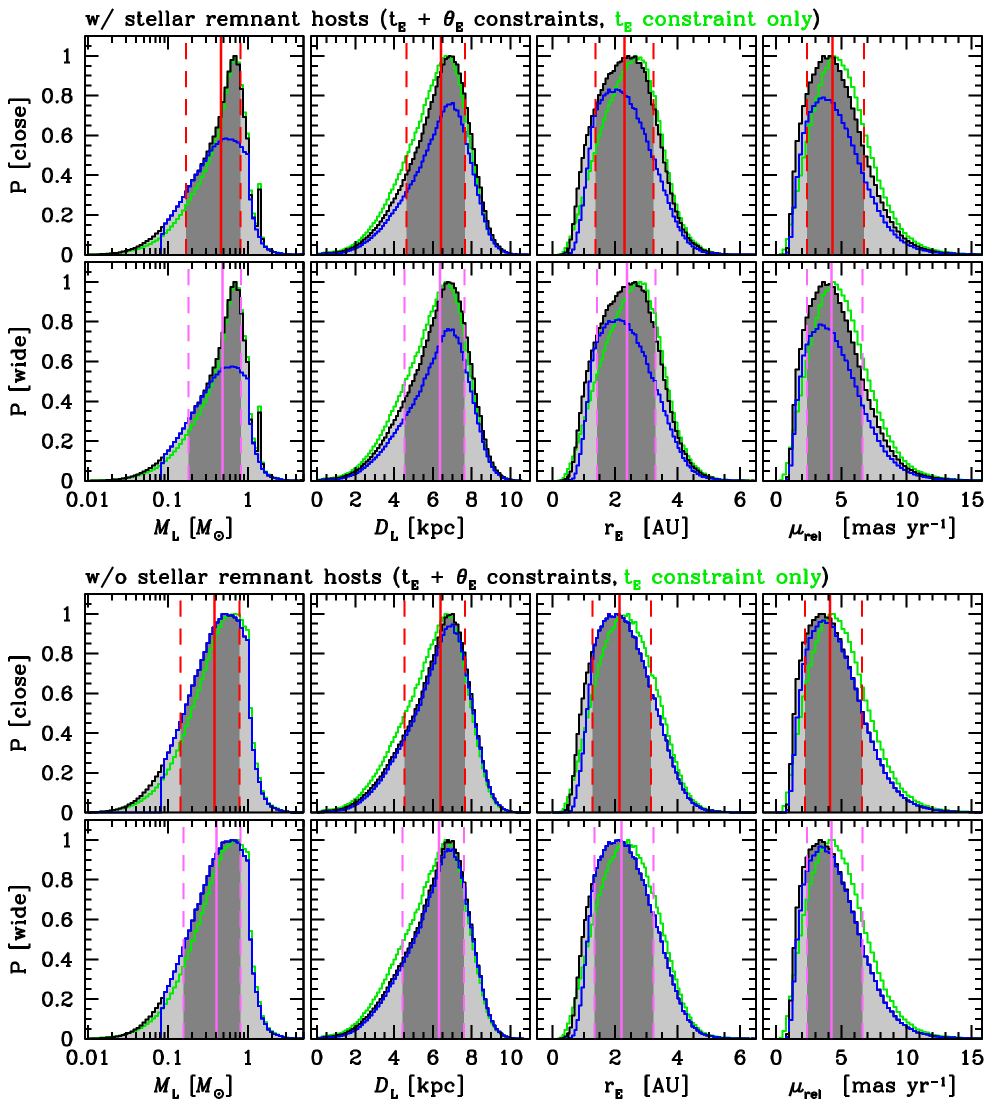}
\caption{
Probability distributions of the lens properties for KMT-2017-BLG-0962. The upper six panels show the probability distributions of the host mass ($M_{\rm L}$), the distance to the lens ($D_{\rm L}$), the physical Einstein ring radius ($r_{\rm E}$), and the lens-source relative proper motion ($\mu_{\rm rel}$) for the close and wide cases. These distributions are constructed from the Galactic prior with stellar remnant hosts. The lower six panels show the probability distributions for the same lens properties, which are constructed from the Galactic prior without stellar remnant hosts. The solid and dashed vertical lines indicate the median value and $68\%$ confidence interval ($1\sigma$ uncertainty) of each property, respectively. The red and pink represent close and wide cases, respectively. The distributions in blue indicate the probability distributions including both the $t_{\rm E}$ and $\theta_{\rm E}$ constraints but considering only luminous hosts. The distributions in green indicate the probability distributions excluding the $\theta_{\rm E}$ constraint.
\label{fig:Bayesian0962}}
\end{figure}

\begin{figure}[htb!]
\epsscale{1.10}
\plotone{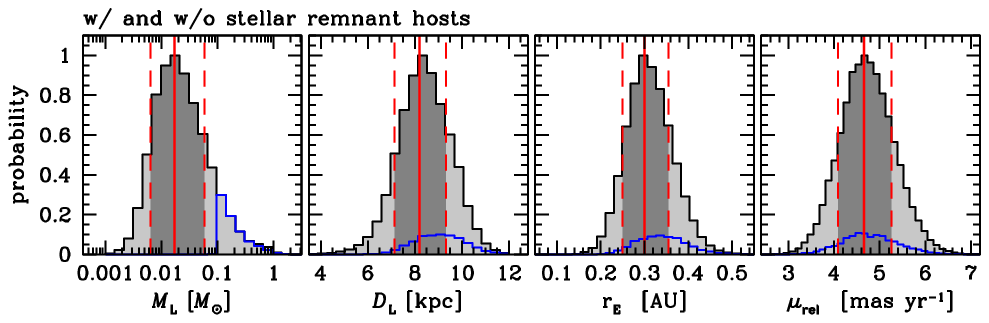}
\caption{
Probability distributions of the lens properties for KMT-2017-BLG-1119. The description is the same as for Figure \ref{fig:Bayesian0962}. In this case, the probability distributions with and without stellar remnant hosts are identical. Thus, we present only one case to avoid clutter.
\label{fig:Bayesian1119}}
\end{figure}

\subsubsection{KMT-2017-BLG-1119}
For this event, MOA {\it R}-band observations exist. Thus, we can measure the ($R-I$) color of the source from source fluxes of the model fits of MOA ($F_{\rm S, MOA}$) and KMTNet ($F_{\rm S, KMT, pyDIA}$) light curves: $(R-I_{\rm KMT})_{\rm S} = -24.684\pm0.021$. Then, we cross-match stars between the KMTNet and MOA CMDs with the OGLE-III catalog \citep{szymanski11} to derive a relation to convert $(R-I_{\rm KMT})$ to $(V-I)_{\rm OGLE-III}$. By combining the measured ($R-I$) source color and conversion relation, we can determine the position of the source on the cross-matched CMD (in OGLE-III magnitude scales): $(V-I, I)_{\rm S,OGLE-III} = (2.425\pm0.105, 19.891\pm0.042)$. Then, by adopting the method of \citet{yoo04} and intrinsic color \citep[$1.06;$][]{bensby11} and magnitude \citep[$14.581;$][]{nataf13} of the red giant clump, we can measure the de-reddened ($V-I$) source color: $(V-I, I)_{\rm S, 0} = (1.060\pm0.105, 18.162\pm0.052)$. Then, we determine the $\theta_{\ast}$ using the color/surface-brightness relation \citep{kervella04}:
\begin{equation}
\theta_{\ast} = 1.093\pm0.131 \, {\mu{\rm as}}.
\end{equation}
In Figure \ref{fig:cmd}, we present the de-reddened KMTNet CMD with positions of the source and centroid of the red giant clump.

\subsection{Bayesian Results}

\subsubsection{KMT-2017-BLG-0962}
 For KMT-2017-BLG-0962, we expect the $\theta_{\rm E}$ constraint (combined with $W(\rho_{\ast})$ and $\theta_{\ast}$) to have only a weak effect on the Bayesian analysis because the constraint of $W(\rho_{\ast})$ is weak for this event (see Figure \ref{fig:dist0962}). In addition, we have had to estimate $\theta_{\ast}$ by estimating the source ($V-I$) using {\it HST} observations of Baade's window rather than making a direct measurement. Thus, we conduct Bayesian analyses with and without the $\theta_{\rm E}$ constraint. In addition, the posterior distributions are constructed using Galactic priors with and without stellar remnants as hosts of the lens system because we cannot rule out the possibility of stellar remnant hosts. Thus, for the degenerate 2L1S solutions (i.e., close and wide), we conduct four types of Bayesian analyses. In Figure \ref{fig:Bayesian0962}, we present the results of the Bayesian analyses. In Table \ref{table:properties}, we present median values of the distributions as representative of the lens system with $68\%$ ($1\sigma$) confidence intervals. The Bayesian results both with and without the $\theta_{\rm E}$ constraint are consistent considering the confidence intervals. The results indicate that this event can be produced by a planetary system consisting of a mid-M dwarf host star and a super Jupiter-mass planet orbiting beyond the snow line.

\subsubsection{KMT-2017-BLG-1119}
 For KMT-2017-BLG-1119, the 2L1S interpretation is disfavored considering the CFHT measurement of the baseline object. Although the 2L1S solution is disfavored, we report the Bayesian results for completeness. In Figure \ref{fig:Bayesian1119}, we also present the probability distributions of the lens properties. Because the timescale of this event is particularly short, the distributions with and without stellar remnant hosts show identical results. Thus, we present one case. In Table \ref{table:properties}, we also present median values of the distributions. 

 The Bayesian results suggest that the lens system of this event may be interesting. If the 2L1S solution is correct, the lens system is most likely to be a sub-Saturn-mass planet with a mass $\sim 0.16\, M_{\rm Jupiter}$ ($\sim 0.53\, M_{\rm Saturn}$) orbiting a brown dwarf host with a mass $\sim 0.017\, M_{\odot}$. Indeed, these kinds of planetary systems having faint/dark hosts ($M_{\rm h } \lesssim 0.08\, M_{\odot}$) were discovered by the microlensing method \citep[e.g.,][]{bennett08, han13, sumi16, shvartzvald17, jung18a, jung18b, miyazaki18}. Microlensing is one useful method to search these kinds of systems because the method can discover planets regardless of the brightness of the hosts. However, we note that the 2L1S interpretation for this event is disfavored. Thus, it is unclear whether or not this event contains an example of such a planetary system.

\section{Conclusion}
 We have presented the analysis of two microlensing events with candidate planets. From Bayesian analysis, we determine the properties of the planet candidates. For KMT-2017-BLG-0962, the lens system may consist of a super Jupiter-mass planet and a mid-M dwarf host. However, the severe 2L1S/1L2S degeneracy of this event, which is unresolvable, prevents claiming this planet discovery with certainty. For KMT-2017-BLG-1119, the 2L1S interpretation would indicate that the lens system consists of a sub-Saturn-mass planet and a brown dwarf host. However, the CFHT imaging supports the 1L2S interpretation rather than this potential interesting planetary system. The planetary solution could be tested with the possibility of conclusively ruling it out by a future measurement of the lens-source relative proper motion.

 The 2L1S/1L2S degeneracies described in this work (and also the degeneracy in \citealt{jung17a}) are far different from the degeneracy for small, short-duration positive anomalies shown in \citet{gaudi98}. The anomalies are of much longer duration and affect a significant fraction of the light curves, yet the degeneracy remains. In addition, the magnitude difference ($\Delta I$) between the two sources is not very extreme ($\Delta I < 1.8$) in contrast to Gaudi's case. These events are similar to the event recently analyzed in \citet{dominik19}. These cases show that the 2L1S/1L2S degeneracy can exist for a wide range of planetary events and for much less extreme binary source systems. Because binary stars are common and this degeneracy has proven not to be limited to a rare subset of binaries, the 2L1S/1L2S degeneracy may be a bigger problem for the discovery of planets than previously thought.

\mbox{}

\acknowledgments 
This research has made use of the KMTNet system operated by the Korea Astronomy and Space Science 
Institute (KASI) and the data were obtained at three host sites of CTIO in Chile, SAAO in South 
Africa, and SSO in Australia. 
Work by IGS and AG was supported by JPL grant 1500811.
AG acknowledges the support from NSF grant AST-1516842.
AG received support from the European Research Council under the European Unions Seventh Framework Programme (FP 7) ERC Grant Agreement n. [321035].
Work by CH was supported by the grant (2017R1A4A1015178) of National Research Foundation of Korea.
This research uses data obtained through the Telescope Access Program (TAP), which has been funded by the National Astronomical Observatories, Chinese Academy of Sciences, and the Special Fund for Astronomy from the Ministry of Finance.
Work by MTP was partially supported by NASA grants NNX16AC62G and NNG16PJ32C.
Work by WZ and PF was supported by Canada-France-Hawaii Telescope (CFHT).
The MOA project is supported by JSPS KAKENHI Grant Number JSPS24253004, JSPS26247023, JSPS23340064, JSPS15H00781, and JP16H06287.

\appendix

\section{Two Parameterizations of the 1L2S Interpretation}

\begin{figure*}[htb!]
\epsscale{1.00}
\plotone{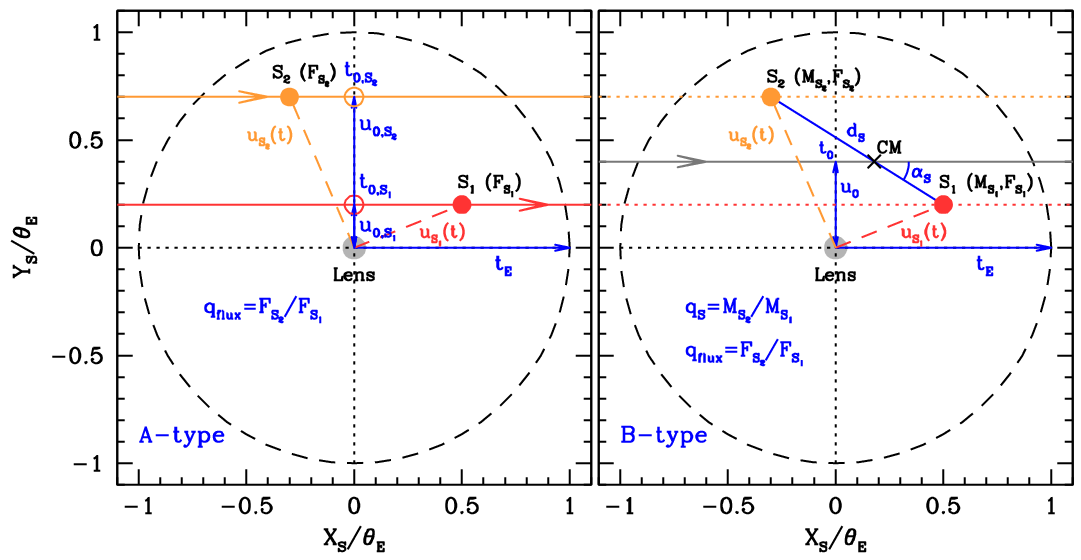}
\caption{
Conceptual geometries of the 1L2S interpretation. The left and right panels present the geometries of 
the A-type and B-type parameterizations, respectively. The blue text indicates parameters. 
The index $i=1$ and $2$ indicate the first source ($S_{1}$) and second source 
($S_{2}$), respectively. The $F_{S_{i}}$ and $M_{S_{i}}$ denote the flux and mass of each source. 
``CM'' denotes the barycenter (i.e., center of mass) of the binary-source system. 
\label{fig:app}}
\end{figure*}

 In Figure \ref{fig:app}, we present conceptual geometries of the 1L2S interpretation for two types of parameterizations. The A-type parameterization (see the left panel of Figure \ref{fig:app}) requires six parameters: $t_{0,S1}$, $t_{0,S2}$, $u_{0,S1}$, $u_{0,S2}$, $t_{\rm E}$, and $q_{\rm flux}$ \citep{griest92}. The first five parameters are directly related to the source trajectories: $t_{0,S_{1}}$ and $t_{0,S_{2}}$ are the time when each source most closely approaches the reference position (i.e., the position of the lens), $u_{0,S_{1}}$ and $u_{0,S_{2}}$ represent the closest separation between each source and the reference position at the time of $t_{0,S_{1}}$ and $t_{0,S_{2}}$, respectively, $t_{\rm E}$ is the Einstein timescale. We use one $t_{\rm E}$ parameter assuming that the lens-source relative speeds are same for both sources, i.e., a comoving binary-source system.  The last parameter, $q_{\rm flux}=F_{S_{2}}/F_{S_{1}}$ is the flux ratio of the sources. The role of $q_{\rm flux}$ is to weight the two 1L1S light curves produced by the individual sources. 

 By adopting this parameterization, the position of each source as a function of time ($t$) is defined in Cartesian coordinates normalized by $\theta_{\rm E}$ as
\begin{equation}
\left[ {\rm X}_{S_{i}}(t), {\rm Y}_{S_{i}}(t) \right] = \left[ \left( \frac{t-t_{0,S_{i}}}{t_{\rm E}} \right), u_{0,S_{i}} \right]~;~ i=1,2.
\end{equation}
According to the positions of the sources, the magnification of each source, ${\rm A}_{S_{i}}(t)$, is defined as
\begin{equation}
{\rm A}_{S_{i}}(t) = \frac{u_{S_{i}}^{2}(t) +2}{u_{S_{i}}(t) \sqrt{u_{S_{i}}^{2}(t) +4}}
~;~
u_{S_{i}}(t) = \left[ {\rm X}_{S_{i}}^{2}(t) + {\rm Y}_{S_{i}}^{2}(t) \right]^{\frac{1}{2}}
~;~
i=1,2~.
\end{equation}
These magnifications are superposed by weighting by the ratio of source fluxes, $q_{\rm flux} = F_{S_{2}}/F_{S_{1}}$. Then, the final magnification of the lensing light curve, ${\rm A}(t)$, is calculated as
\begin{equation}
{\rm A}(t) = \frac{{\rm A}_{S_{1}}(t) + q_{\rm flux}\,{\rm A}_{S_{2}}(t)}{1+q_{\rm flux} }.
\end{equation}
This model light curve in the magnification scale is converted to the flux scale of each dataset for comparison to the observations using two additional parameters, $F_{S}$ and $F_{B}$ (similar to those of the 2L1S interpretation). These additional parameters are determined using the least-square fitting method.

 The merit of this A-type parameterization is that it is possible to directly guess the initial values of most parameters (except $q_{\rm flux}$) from the observed light curve. However, the A-type parameterization has a disadvantage that it is difficult to apply higher-order effects, especially the orbital motion of the binary-source system.

 Thus, we introduce an alternative parameterization, B-type (see right panel of Figure \ref{fig:app}), which considers the motion of the barycenter of the binary-source system \citep{jung17a}, rather than the motion of each source. To describe the barycenter motion, it requires three parameters ($t_0$, $u_0$, $t_{\rm E}$): $t_{0}$ is the time when the barycenter closely approaches to the reference position, $u_{0}$ is the closest separation at the time of $t_{0}$, and $t_{\rm E}$ is the Einstein timescale. To derive the trajectory of each source from the barycenter trajectory, three additional parameters ($d_{S}$, $q_{S}$, $\alpha_{S}$) are required to describe the binary-source system: $d_{S}$ is the projected separation between the sources, $q_{S}=M_{S_{2}}/M_{S_{1}}$ is a mass ratio of the source stars, and $\alpha_{S}$ is an angle between the axis of the binary-source and the barycenter trajectory. In addition, there is the last parameter ($q_{\rm flux}$) that is identical to that of the A-type parameterization.

By adopting this parameterization, the source positions are defined as
\begin{equation}
\begin{bmatrix}
{\rm X}_{S_{i}}(t) \\ 
{\rm Y}_{S_{i}}(t) \\
\end{bmatrix}
=
\begin{bmatrix}
{\rm X}_{\rm CM}(t) \pm {\rm r}_{S_{i}} \cos{\alpha_{S}}  \\
{\rm Y}_{\rm CM}(t) \mp {\rm r}_{S_{i}} \sin{\alpha_{S}}  \\
\end{bmatrix}
~;~
\left[ {\rm X}_{\rm CM}(t), {\rm Y}_{\rm CM}(t) \right] = \left[ \left( \frac{t-t_{0}}{t_{\rm E}} \right), u_{0} \right]
~;~
i=1,2~,
\end{equation}
where the $r_{S_{1}}$ and $r_{S_{2}}$ are the separations between the barycenter and each source, which are defined as
\begin{equation}
{\rm r}_{S_{1}} = d_{S} \left( \frac{q_{S}}{1+q_{S}} \right)
~;~
{\rm r}_{S_{2}} = d_{S} \left( \frac{1}{1+q_{S}} \right).
\end{equation}
Based on the positions of the source, the final model light curve is constructed in the same way as the previous parameterization (see Equations A2 and A3).

 This B-type parameterization has merit when higher-order effects are considered. In particular, the orbital motion of the binary-source can be easily introduced because the binary source positions are defined from the barycenter. To introduce the source-orbital motion, two additional parameters, $d{d_{S}}/dt$ and $d{\alpha_{S}}/dt$, are required. These parameters are the variation rates of $d_{S}$ and $\alpha_{S}$ to describe a partial orbit of the binary-source system. The variations are derived as
\begin{equation}
d'_{S} = d_{S} + \frac{d{d_{S}}}{dt} \left( t-t_{\rm ref} \right)
~;~
\alpha'_{S} = \alpha_{S} + \frac{d{\alpha_{S}}}{dt} \left( t-t_{\rm ref} \right),
\end{equation}
where $t_{\rm ref}$ is a reference time for describing the orbital motion of sources (we set $t_{\rm ref}=t_{0}$ for the modeling in this work).
Thus, the source trajectories are varied by the source-orbital motion, which are described by modifying Equations A4 and A5 as,
\begin{equation}
\begin{bmatrix}
{\rm X'}_{S_{i}}(t) \\ 
{\rm Y'}_{S_{i}}(t) \\
\end{bmatrix}
=
\begin{bmatrix}
{\rm X}_{\rm CM}(t) \pm {\rm r'}_{S_{i}} \cos{{\alpha'}_{S}}  \\
{\rm Y}_{\rm CM}(t) \mp {\rm r'}_{S_{i}} \sin{{\alpha'}_{S}}  \\
\end{bmatrix}
~{\rm where}~
{r'}_{S_{1}} = d'_{S} \left( \frac{q_{S}}{1+q_{S}} \right)
~;~
{r'}_{S_{2}} = d'_{S} \left( \frac{1}{1+q_{S}} \right).
\end{equation}

 However, the downside of this B-type parameterization is that it is particularly difficult to guess the initial parameters for describing the binary-source system (i.e., $d_{S}$, $q_{S}$, and $\alpha_{S}$). Thus, usually, this parameterization is only adopted for testing higher-order effects.


\section{Non-detections of Higher-order Effects}

\subsection{The Annual Microlens Parallax (APRX) Effect of the 2L1S Interpretation}

\begin{figure}[htb!]
\epsscale{0.95}
\plotone{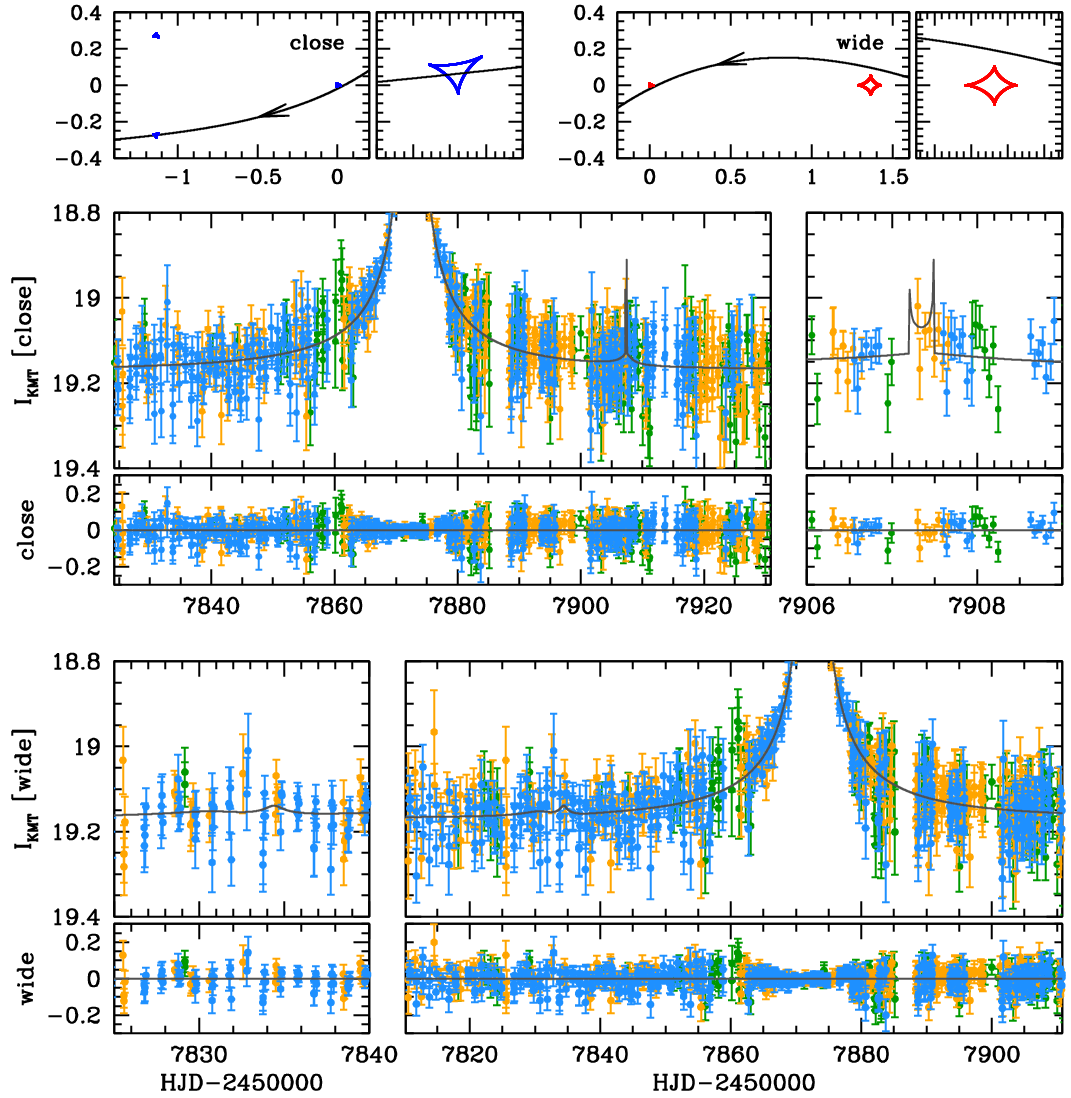}
\caption{
APRX models (2L1S) of KMT-2017-BLG-0962. Upper panels show geometries of the APRX models for the close (left) and wide (right) cases with zoom-ins where caustic-crossing and approach. Middle panels show the APRX model light curve (solid line) of the close case with a zoom-in where the part of caustic-crossing (left panels). Bottom panels show the APRX model light curve of the wide case. The zoom-in (right) shows the light curve part where the caustic approach. Bottom panels of each light curve show residuals between models and observations. The color scheme of the observations is identical to Figure \ref{fig:lc0962}.
\label{fig:aprx0962}}
\end{figure}

 The APRX is caused by the orbital motion of Earth \citep{gould92}. Thus, the Einstein timescale ($t_{\rm E}$) is a direct indicator for estimating the possibility of detecting the APRX signal. Empirically, to detect the APRX signal, the event should last more than $\sim20$ days. For KMT-2017-BLG-0962, $t_{\rm E}$ is about $33$ days, which implies that there is a chance to detect the APRX signal in the light curve. Thus, we try to measure the APRX by introducing two additional parameters, $\pi_{{\rm E},N}$ and $\pi_{{\rm E},E}$, which indicate the north and east directions of the microlens parallax vector ($\pivec_{\rm E}$), respectively. From the model considering the APRX, we find $\chi^{2}$ improvements, $13.0$ and $0.1$, for the close and wide cases, respectively. However, these improvements originate in fits of systematics in the baseline, which are caused by accidental caustic-crossing and approach (See Figure \ref{fig:aprx0962}). This fact implies that the APRX is not significantly constrained in these fits. Thus, we cannot extract any useful information from the APRX model for this event. For KMT-2017-BLG-1119, $t_{\rm E}$ is only $2.9$ days, which means that it is unlikely to exist the APRX signal can be detected in the light curve. However, for consistency, we also test the APRX model for this event. From the model, as expected, the APRX signal is not detected.


\subsection{The Source-orbital Effect of the 1L2S Interpretation}

\begin{figure}[htb!]
\epsscale{0.95}
\plotone{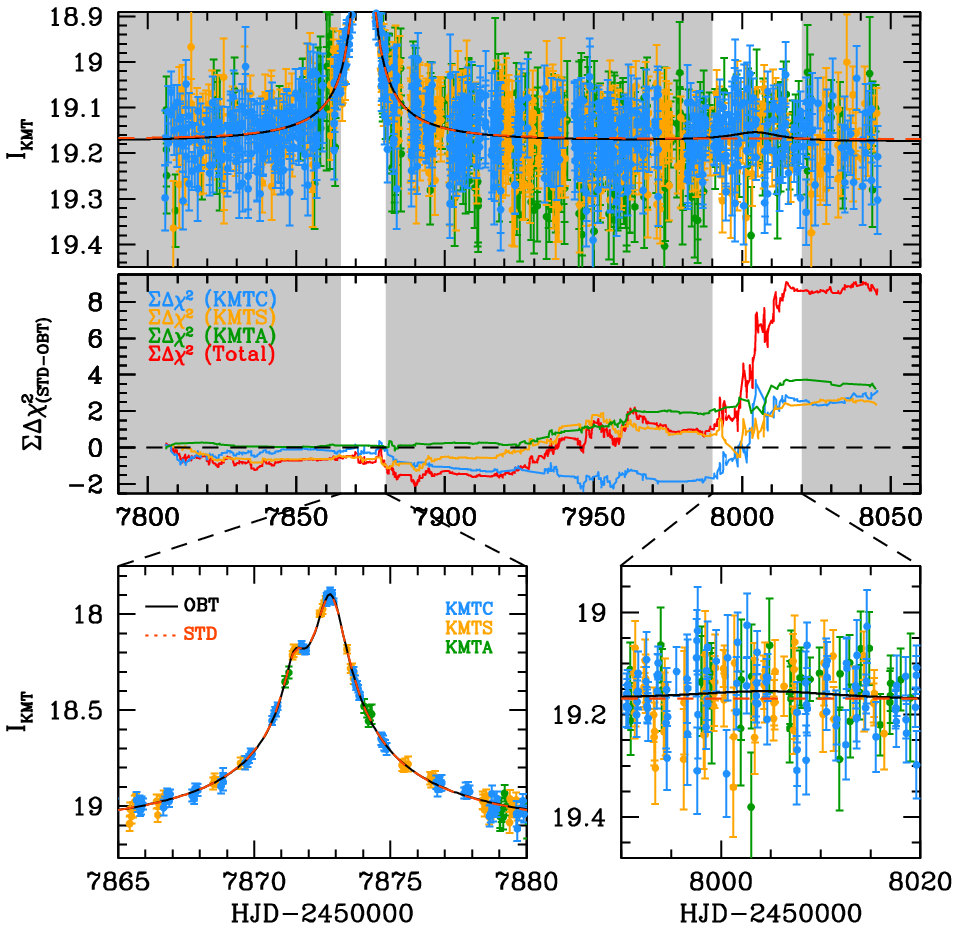}
\caption{
Cumulative $\chi^2$ difference ($\Sigma\Delta\chi^2$) between static and source-orbital models (1L2S) of KMT-2017-BLG-0962. The upper panel shows the whole baseline observations with static (dashed line in scarlet) and source-orbital (solid line in black) model light curves. The bottom panel presents the $\Sigma\Delta\chi^2$ of each dataset. The boxes in white show light curve zoom-ins for anomaly part (left) and the perturbation induced by the source-orbital motion effect (right).
\label{fig:obt0962}}
\end{figure}

 For the 1L2S interpretation, the binary sources always orbit each other to conserve their angular momentum. As a result, this source-orbital motion can affect the light curve if the microlensing event was caused by the 1L2S. It implies that once we may find the source-orbital effect on the lensing light curve, we can obtain a key clue to resolve the 2L1S/1L2S degeneracy. Therefore, we test the effect by introducing additional parameters of the simplified source-orbital motion (see Appendix A, B-type parameterization). The possibility of the detection of the source-orbital effect depends on the timescale of the event, similar to the APRX effect. As expected, for KMT-2017-BLG-1119, there is no $\chi^2$ improvement considering the very short $t_{\rm E}$ of this event. In contrast, for KMT-2017-BLG-0962, we find a small $\chi^2$ improvement ($\Delta\chi^2\sim8.7$) when the source-orbital effect is considered. We investigate this improvement using the cumulative $\chi^2$ difference plot. See Figure \ref{fig:obt0962}. From the investigation, we find that the improvement mostly comes from the fitting of the ``bump-like'' feature in the baseline (${\rm HJD'}\sim8000$). It is unclear whether this feature is real or due to some systematics in the baseline of the event. With $\Delta\chi^{2}\sim8.7$ for $2$ additional degree of freedom, the significance is too low to claim a detection.



\end{document}